\documentclass[%
 reprint,
superscriptaddress,
 amsmath,amssymb,
 aps,
pra,
]{revtex4-2}
\usepackage[dvipsnames]{xcolor}
\usepackage[english]{babel}
 
\makeatletter
\def\bbl@set@language#1{%
  \edef\languagename{%
    \ifnum\escapechar=\expandafter`\string#1\@empty
    \else\string#1\@empty\fi}%
  \@ifundefined{babel@language@alias@\languagename}{}{%
    \edef\languagename{\@nameuse{babel@language@alias@\languagename}}%
  }%
  \select@language{\languagename}%
  \expandafter\ifx\csname date\languagename\endcsname\relax\else
    \if@filesw
      \protected@write\@auxout{}{\string\select@language{\languagename}}%
      \bbl@for\bbl@tempa\BabelContentsFiles{%
        \addtocontents{\bbl@tempa}{\xstring\select@language{\languagename}}}%
      \bbl@usehooks{write}{}%
    \fi
  \fi}
\newcommand{\DeclareLanguageAlias}[2]{%
  \global\@namedef{babel@language@alias@#1}{#2}%
}
\makeatother

\DeclareLanguageAlias{en}{english}

\usepackage{graphicx}% Include figure files
\usepackage{amsmath}
\usepackage{array}
\usepackage{amssymb}
\usepackage{dcolumn}% Align table columns on decimal point
\usepackage{bm}% bold math
\let\savecorresponds\corresponds
\let\corresponds\relax
\usepackage{mathabx}
\usepackage{enumitem} 
\usepackage{tabularx}
\usepackage{soul}
\usepackage{ulem}
\let\corresponds\savecorresponds
\usepackage{hyperref}% add hypertext capabilities
\usepackage[utf8]{inputenc}
\renewcommand{\vec}[1]{\mathbf{#1}}

\newcommand{\brakket}[3]{\ensuremath{\langle{#1}|{#2}|{#3}\rangle}}
\def\ket#1{ | #1 \rangle}
\def\bra#1{{\langle #1 |  }}
\def\vec#1{\boldsymbol{#1}}

\def\pd2v#1#2#3{\frac{\partial^2 #1}{\partial #2 \partial #3}}

\def \vec#1{\mathbf{#1}}
\def \2x2mat#1#2#3#4{
\left( \begin{array}{cc}
#1 &  #2 \\  #3 &  #4
\end{array} \right)
}

\begin{document}

\preprint{APS/123-QED}

\title{Manipulation and certification of high-dimensional entanglement through a scattering medium}%  \\

\author{Baptiste Courme}
\affiliation{Laboratoire Kastler Brossel, ENS-Université PSL, CNRS, Sorbonne Université, Collège de France, 24 rue Lhomond, 75005 Paris, France\
}%
\author{Patrick Cameron}
\affiliation{School of Physics and Astronomy, University of Glasgow, Glasgow G12 8QQ, UK\
}%
\author{Daniele Faccio}%
\affiliation{School of Physics and Astronomy, University of Glasgow, Glasgow G12 8QQ, UK\
}%
\author{Sylvain Gigan}
\affiliation{Laboratoire Kastler Brossel, ENS-Université PSL, CNRS, Sorbonne Université, Collège de France, 24 rue Lhomond, 75005 Paris, France\
}%
\author{Hugo Defienne} \email[Corresponding author: ]{hugo.defienne@insp.upmc.fr}
\affiliation{School of Physics and Astronomy, University of Glasgow, Glasgow G12 8QQ, UK\ 
}
\affiliation{Sorbonne Université, CNRS, Institut des NanoSciences de Paris, INSP, F-75005 Paris, France\
}

\date{\today}
\begin{abstract}
High-dimensional entangled quantum states improve the performance of quantum technologies compared to qubit-based approaches. In particular, they enable quantum communications with higher information capacities or enhanced imaging protocols. However, the presence of optical disorder such as atmospheric turbulence or biological tissue perturb quantum state propagation and hinder their practical use. Here, we demonstrate a wavefront shaping approach to transmit high-dimensional spatially-entangled photon pairs through scattering media. Using a transmission matrix approach, we perform wavefront correction in the classical domain using an intense classical beam as a beacon to compensate for the disturbances suffered by a co-propagating beam of entangled photons. Through violation of an Einstein-Podolski-Rosen criterion by $988$ sigma, we show the presence of entanglement after the medium. Furthermore, we certify an entanglement dimensionality of $17$. This work paves the way towards manipulation and transport of entanglement through scattering media, with potential applications in quantum microscopy and quantum key distribution.
\end{abstract}

\maketitle

{\section{Introduction}}

Quantum entanglement plays a central role in quantum technologies. In this respect, high-dimensional entangled states of light offer higher information capacities~\cite{mirhosseini_high-dimensional_2015} and better resistance to noise~\cite{ecker_overcoming_2019} over qubit-based quantum communication systems. In particular, their high tolerance to losses make them good candidates for the realization of device-independent quantum communication~\cite{acin_device-independent_2007}. Furthermore, they also serve as an essential resource in many quantum imaging protocols, including sub-shot-noise imaging~\cite{brida_experimental_2010}, resolution and sensitivity-enhanced approaches~\cite{toninelli_resolution-enhanced_2019,camphausen_quantum-enhanced_2021,defienne_pixel_2022-2,ndagano_quantum_2022}, quantum illumination~\cite{defienne_quantum_2019-1,gregory_imaging_2020} and quantum holography~\cite{devaux_quantum_2019,defienne_polarization_2021-4}.

An important issue to be overcome in these applications is the preservation of entanglement after transmission through optical disorder. Light scattering in biological tissue, atmospheric turbulence, random mode mixing in multimode fibers, are examples of adverse effects that can significantly impair the performance of imaging and communication systems. In classical optics, wavefront shaping techniques~\cite{vellekoop_focusing_2007,popoff_measuring_2010} were developed to mitigate these effects. Such an ability to control light propagation through scattering samples has led to many technological advances, such as the transmission of spatial information through multimode fibers~\cite{carpenter_110x110_2014,ploschner_seeing_2015-3} or deep tissue imaging~\cite{kang_imaging_2015,badon_smart_2016}.
\begin{figure*}
\includegraphics[width=0.9 \textwidth]{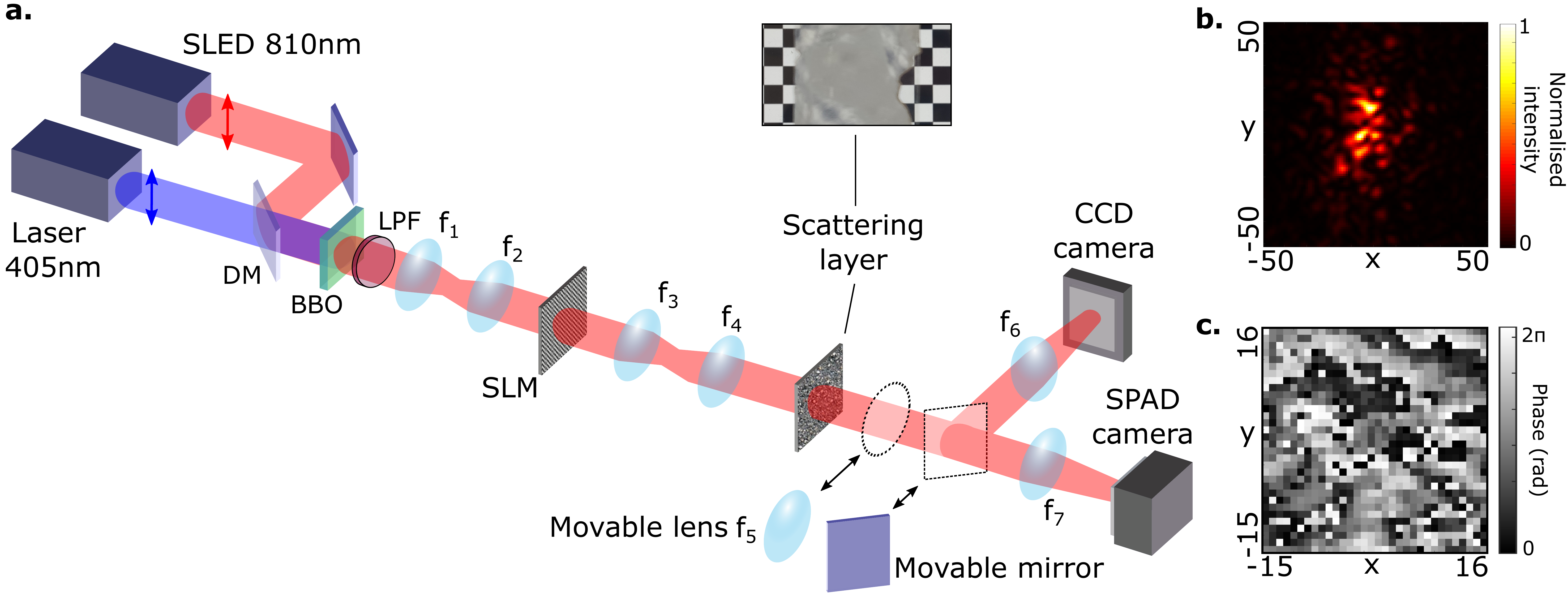} % this command will be ignored
\caption{\label{Figure1} \textbf{Experimental setup.} \textbf{a,} Spatially-entangled photon pairs are produced by type-I spontaneous parametric down-conversion (SPDC) by illuminating a $\beta$-Barium Borate (BBO) crystal (0.5mm thickness) with a vertically polarized collimated laser diode ($405$nm). Simultaneously, horizontally polarized and collimated light emitted by a super luminescent diode (SLED) is aligned along the pump beam using a dichroic mirror(DM). Long-pass and band-pass filters at $810 \pm 5$nm (LPF) remove pump photons and filters the classical light. A two-lenses system $f_1-f_2$ images the crystal surface onto onto a spatial light modulator (SLM), that is itself imaged by $f_3-f_4$ onto a scattering medium (layer of parafilm on a microscope slide). In the momentum-basis configuration, a single-lens Fourier imaging system ($f_6$ or $f_7$) is used to image output light either onto a single-photon avalanche diode (SPAD) camera (with the movable mirror) or onto a Charge Coupled Device (CCD) camera (without the movable mirror). In the position-basis configuration, a movable lens $f_5$ is inserted to image the scattering layer onto the cameras. \textbf{b,} Intensity image acquired by the CCD camera using classical light and no SLM correction. \textbf{c,} Correction SLM pattern calculated using the transmission matrix. Note that the SLM is represented in transmission while it operates in reflection. Spatial units are in pixels.}
\end{figure*}

Wavefront shaping has also been used to manipulate non-classical light through scattering media, such as single~\cite{carpenter_mode_2013,huisman_controlling_2014,defienne_nonclassical_2014} and indistinguishable photons~\cite{defienne_two-photon_2016,wolterink_programmable_2016}. These techniques have recently been applied to high-dimensional spatially-entangled photon pairs. Proof-of-principle experiments include their transmission through thin static~\cite{defienne_adaptive_2018-3,devaux_restoring_2022} and dynamic diffusers~\cite{lib_real-time_2020-1}, and multimode fibers~\cite{valencia_unscrambling_2020-1}. However, these demonstrations have several limits. In references~\cite{defienne_adaptive_2018-3,lib_real-time_2020-1,devaux_restoring_2022}, the authors demonstrated only the recovery of spatial correlations after the medium by performing measurements in a single spatial basis, but not entanglement itself. In reference~\cite{valencia_unscrambling_2020-1}, entanglement was certified at the output up to six dimensions. However, the proposed method works exclusively if the output state is characterized through series of single-outcome spatial-mode measurements. In such case, the number of measurements scales as $2d^2$, where $d$ is the local dimension of the system, which leads to very long acquisition times that are often prohibitive for real-world applications. More importantly, single-outcome  measurement schemes use the assumption that the total number of coincidences in all modes does not fluctuate from one single-outcome measurement to the next, which is wrong in general~\cite{friis_entanglement_2019}. Such an assumption would not be acceptable in an adversarial scenario such as quantum key distribution as it will compromise the security of the protocol.

Here, we demonstrate the transport of high-dimensional entanglement through a scattering medium using a transmission-matrix-based wavefont shaping approach and a multi-outcome spatial measurement method. The transmission matrix of the scattering medium is first measured using classical light~\cite{popoff_measuring_2010}. Then, a spatial light modulator (SLM) is programmed with a correction phase mask to mitigate scattering of entangled photons. Photon correlation measurements in position and momentum are performed at the output using a commercially available single-photon avalanche diode (SPAD) camera~\cite{ndagano_imaging_2020-1}. The presence of entanglement after the medium is demonstrated through violation of an Einstein-Podolski-Rosen (EPR) criterion~\cite{einstein_can_1935,giovannetti_characterizing_2003} with a confidence of $988$ sigma. From our measurements in position and momentum bases, we finally certify high-dimensional entanglement in up to $17$ dimensions, using the method developed by Bavaresco et al.~\cite{bavaresco_measurements_2018}.\\

{\section{Experimental setup}}

Spatially-entangled photon pairs are produced via spontaneous parametric down conversion (SPDC) in a thin $\beta$-barium-borate (BBO) crystal using the apparatus shown in Figure~\ref{Figure1}.a. Two $4f$ imaging systems conjugate the output surface of the crystal onto an SLM and then onto a scattering medium. The scattering medium is a stretched layer of parafilm placed on a microscope slide. At the output, photon pairs are detected using a commercial SPAD camera~\cite{bronzi_100_2014}. To measure momentum correlations, we used the lens $f_7$ positioned in a 2\textit{f}-arrangement to image the Fourier plane of the scattering layer onto the camera. To measure position correlations, a lens $f_5$ is inserted to form a 4\textit{f}-arrangement and image the output surface of the scattering layer onto the camera. These two optical configurations allow measurements to be made in both the position and momentum bases, which form two mutually unbiased bases~\cite{erker_quantifying_2017}. In the same apparatus, light emitted by a super-luminescent diode (SLED) is collimated and superimposed onto the pump beam propagation axis. This classical source is used to measure the transmission matrix $T$ of the scattering medium using the method developed by Popoff et al.~\cite{popoff_measuring_2010}. $T$ was measured between $32 \times 32$ macro-pixels of the SLM (input modes basis) and $192 \times 192$ pixels of a CCD camera (output modes basis) positioned in a Fourier plane of the medium. Figure~\ref{Figure1}.b shows the intensity speckle pattern measured by the CCD with no phase pattern programmed on the SLM. 

Using the transmission matrix formalism, propagation of spatially-entangled photon pairs in our optical system is written as
\begin{equation}
\label{Eq0}
\Psi^{out} = T D \Psi^{in} D^t T^t,
\end{equation}
where $\Psi^{in}$ and $\Psi^{out}$ are the spatial two-photon wave-functions written in the discrete input and output basis, respectively, and $D$ is a diagonal matrix associated with the SLM. $D$ contains $1024$ complex phase terms $\{ e^{i \theta_k} \}_{k \in [\![ 1,1024 ]\!]}$, where $\theta_k$ is the phase term associated with the $k^{th}$ SLM macro-pixel. At this point, it is important to clarify the different mode bases under consideration. At the input, the basis is a position basis where each mode correspond to a SLM macro-pixel. In our experiment, since SLM macro-pixels are much wider ($120 \mu$m) than the position correlation width of entangled photons in the SLM plane ($24 \mu$m), the input two-photon wavefunction can be approximate as $\Psi^{in} \approx 1\!\!1$ (see Appendix G).  At the output, the basis can be a position or a momentum basis according to the chosen configuration (Fig.~\ref{Figure1}.a), where each mode corresponds to a camera pixel. When measuring the transmission matrix, the camera is positioned in a Fourier plane of the system, which means that the output basis is a momentum basis. To switch to a position basis, we insert the lens $f_5$ in the system and substitute $T \rightarrow FT$ in equation~\eqref{Eq0}, where $F$ is a matrix associated with a discrete Fourier transform (see Appendix G). 

To recover entanglement after the medium, one can program the SLM to restore spatial correlations between photon pairs simultaneously in the position and momentum bases, as would be the case without a scattering medium~\cite{ndagano_imaging_2020-1}. In other words, we are looking for a set of phases $\{\theta_k \}_{k \in [\![1,1024]\!]}$ so that $D$ maximizes the following conditions:
\begin{align}
&|T D^2 T^t|^2 = 1\!\!1, \label{C1} \\
&|F T D^2 T^t F^t|^2 = 1\!\!1. \label{C2}
\end{align}
In general, these conditions are difficult to satisfy by shaping the light with a single SLM. In our experiment, however, they can be simplified. First, since the scattering medium is a thin parafilm layer, the term $FT$ is quasi-diagonal, thus satisfying condition~\eqref{C2}. Second, we showed in our simulations that the set of phases $\theta_k = \arg(T_{pk}^*)$, where $p$ is an index associated with the central pixel of the CCD camera, is a solution that partially satisfies condition~\eqref{C1} (see Appendix H). Figure~\ref{Figure1}.c shows the corresponding SLM phase pattern. Note that this solution is the same as the one used to refocus classical light through the medium on the central camera pixel~\cite{popoff_measuring_2010}, through phase-conjugation.\\ 

{\section{Violation of an EPR criterion}}

\begin{figure}
\includegraphics[width=1 \columnwidth]{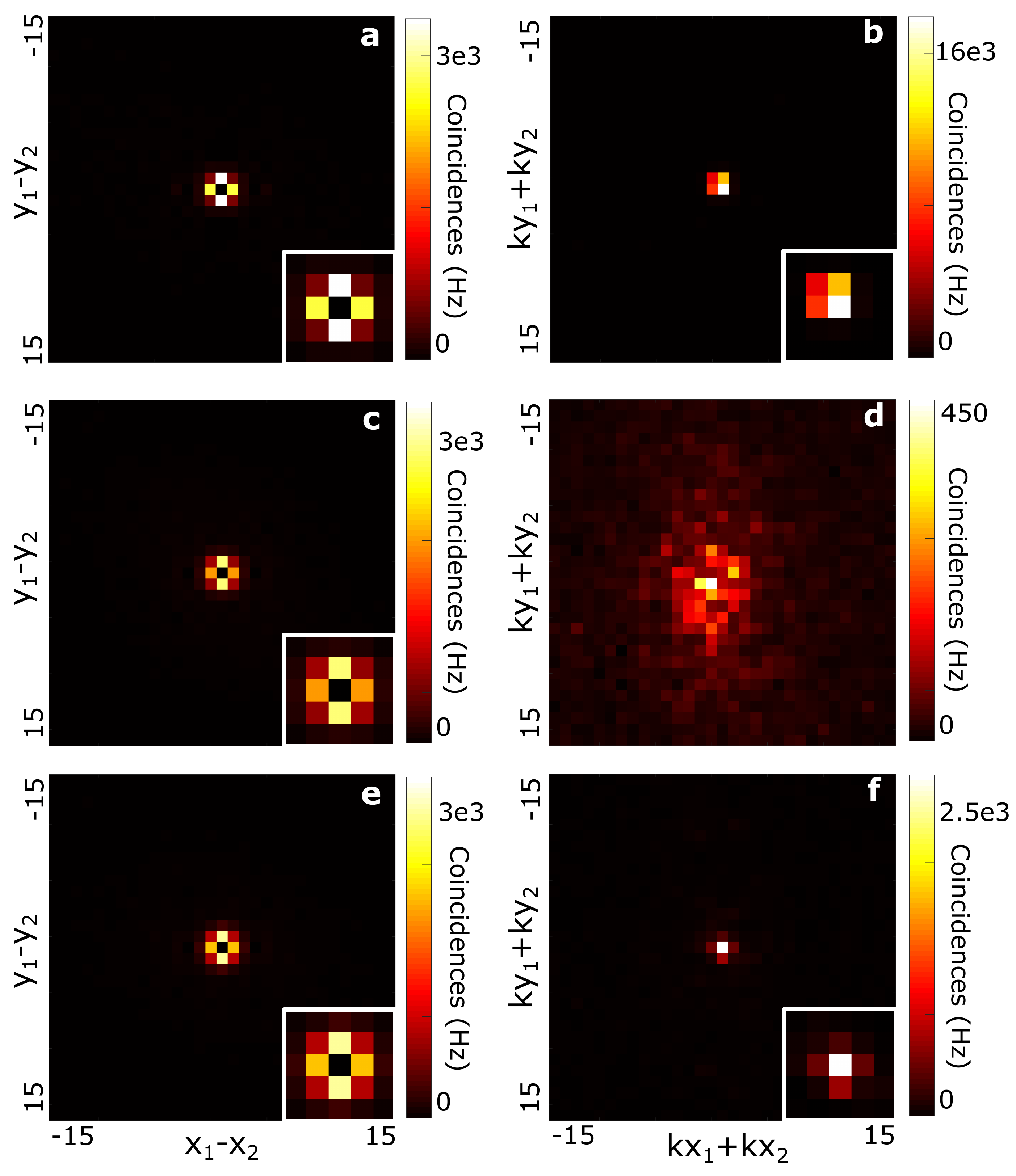} % this command will be ignored
\caption{\label{Figure2} \textbf{EPR criterion violation.} Images of JPD minus-coordinate projections acquired using the position basis configuration without the scattering medium (\textbf{a}), with the medium and no SLM correction (\textbf{c}) and with the medium and SLM correction (\textbf{e}). Images of JPD sum-coordinate projections acquired using the momentum basis configuration without the scattering medium (\textbf{b}), with the medium and no SLM correction (\textbf{d}) and with the medium and SLM correction (\textbf{f}). Analysis was performed on a total of $1.5 \times 10^9$ images. The central pixel in the minus-coordinate projections has been set to zero because the SPAD camera does not resolve the number of photons and therefore cannot measure photon coincidences at the same pixel. Spatial units are in pixels.}
\end{figure}

We demonstrate the presence of spatial entanglement at the output by violating a separability criterion derived by Giovanetti et al.~\cite{giovannetti_characterizing_2003}. It states that separable systems satisfy the joint uncertainty product:
\begin{equation}
\label{Eq1}
\Delta \vec{r} \Delta \vec{k} > \frac{1}{2},
\end{equation}
where the uncertainties $\Delta \vec{r} = \Delta (\vec{r_1}-\vec{r_2})$ and $\Delta \vec{k} = \Delta (\vec{k_1}+\vec{k_2})$ correspond to measures of the correlation widths associated with the joint probability distributions (JPDs) of photon pairs measured in position and momentum, respectively, for pairs of photons labeled 1 and 2. This criterion is commonly used to demonstrate the presence of entanglement in bipartite quantum systems~\cite{howell_realization_2004,moreau_realization_2012,edgar_imaging_2012,dabrowski_einsteinpodolskyrosen_2017}. To estimate $\Delta \vec{r}$ and $\Delta \vec{k}$, we measure the spatial JPD of photon pairs using the SPAD camera in the two configurations described in Figure~\ref{Figure1}~\cite{defienne_general_2018,ndagano_imaging_2020-1}. In the momentum basis configuration, the width of the central peak in the JPD sum-coordinate projection corresponds to $\Delta \textbf{k}$. Accounting for the effective magnification of our optical system, we measured $\Delta \textbf{k} = 1.495(1) \times 10^4$ m$^{-1}$ without the scattering medium (Fig.~\ref{Figure2}.a) and $\Delta \textbf{k} = 1.72(1) \times 10^4$ m$^{-1}$ with the scattering medium and SLM correction (Fig.~\ref{Figure2}.e). Width values are estimated using a Gaussian model~\cite{fedorov_gaussian_2009} (see Appendix D). In the presence of the medium but without SLM correction, spatial correlations become spread and distorted (Fig~\ref{Figure2}.c), making it difficult to properly estimate $\Delta \textbf{k}$. As discussed in reference~\cite{gnatiessoro_imaging_2019-1}, an approximate value can nevertheless be obtained by measuring the width of the envelop $\Delta \textbf{k} = 9.8(1) \times 10^4$ m$^{-1}$.

\begin{figure*}
\includegraphics[width=0.9 \textwidth]{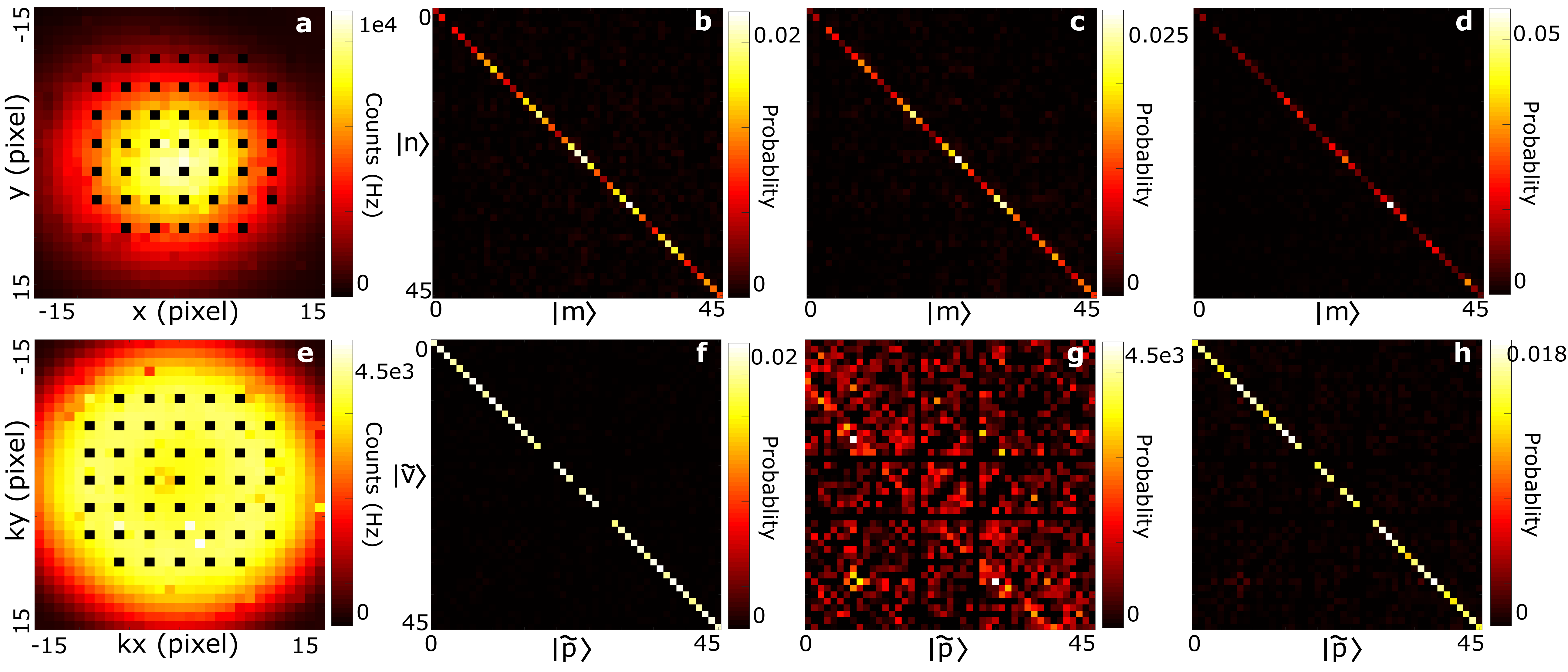} % this command will be ignored
\caption{\label{Figure3} \textbf{Entanglement certification.} \textbf{a,} Intensity image acquired in the position basis configuration showing the grid of $45$ pixels. \textbf{b,c,d,} Correlations matrices between all the pixels in the grid of the position basis without medium \textbf{(b)}, with medium and flat SLM \textbf{(c)} and with medium and correction SLM \textbf{(d)}. Each pixel is labeling the spatial coordinates of photon: $\ket{m}$ in the position basis. \textbf{e,} Intensity image acquired in the momentum basis configuration showing the grid of $45$ pixels. \textbf{f,g,h,} Correlations matrix between all the pixels in the grid in the momentum basis without medium \textbf{(f)}, with medium and flat SLM \textbf{(g)} and with medium and correction SLM \textbf{(h)}. Each pixel is labelling the spatial coordinates of photon: $\ket{\widetilde{p}}$ in the momentum basis. Analysis was performed on a total of $1.5 \times 10^9$ images. Black lines and columns are associated with hot pixels of the sensor set to zero.}
\end{figure*}

We repeated the above analysis in the position basis configuration to extract $\Delta \vec{r}$. Projections of the JPD along the minus-coordinate are measured without the medium (Fig.~\ref{Figure3}.b), with the medium and no SLM correction (Fig.~\ref{Figure3}.d), and with the medium and SLM correction (Fig.~\ref{Figure3}.f). In all cases, a coincidence peak is observed at the centre of the projection that demonstrates strong position correlations between photon pairs. Accounting for the optical magnification, we measured $\Delta \vec{r} = 6.77(1) \times 10^{-6}$ m (no medium), $\Delta \vec{r} = 8.32(2) \times 10^{-6}$ m (medium and no SLM correction) and $\Delta \vec{r} = 8.82(2) \times 10^{-6}$ m (medium and SLM correction) by fitting with a Gaussian (see Appendix D).
The measured values of transverse position and momentum correlation widths violate the separability criterion $\Delta \textbf{r} \Delta \textbf{k} = 0.1013(1) < 1/2$ without medium and $\Delta \textbf{r} \Delta \textbf{k} = 0.1519(2) < 1/2 $ with medium and SLM correction, thus demonstrating the presence of spatial entanglement. Confidence levels are $C=2589$ and $C=988$, respectively (see Appendix E). In the presence of the medium and without SLM correction, a similar calculation shows that $\Delta \textbf{r} \Delta \textbf{k} = 0.82(1) > 1/2$, with confidence level of $C=41$, concluding that the separability criterion is not violated. However, this last calculation must be interpreted with caution because measuring the width of the envelop in Figure~\ref{Figure2}.d to evaluate $\Delta \textbf{k}$ is only approximate, as the latter does not necessarily follow a Gaussian shape. The impossibility to detect the presence of entanglement in this case will be confirmed properly using the certification method in the following part.   \\

{\section{Certification of high-dimensional entanglement}}

Measurements in two mutually unbiased bases (MUBs) enable the use of a recently developed entanglement witness for certifying high-dimensional entanglement~\cite{bavaresco_measurements_2018}. The discrete position and momentum bases can be used as two MUBs and are accessible in our experimental setup~\cite{erker_quantifying_2017}. As shown in Figures~\ref{Figure3}.a and e, we selected $d = 45$ pixels uniformly distributed over a central region in both configurations. Modes associated with the chosen pixels are denoted $\{ \ket{m} \}_{m \in [\![0,d-1]\!]}$ (discrete position basis) and $\{ \ket{\widetilde{p}} \}_{p \in [\![0,d-1]\!]}$ (discrete momentum basis). To certify entanglement dimensionality of the detected state $\rho$, correlation measurements are performed in the two MUBs to compute a lower bound for the fidelity of the state with respect to a maximally entangled target state $\ket{\Psi} = \frac{1}{\sqrt{d}} \sum^{d-1}_{m=0} \ket{m m}$. Without the scattering medium, we obtained a lower
bound value of the fidelity $\widetilde{F}(\rho,\Psi)= 0.6138$ from the correlation matrices measured in Figures~\ref{Figure3}.b and f. The entanglement dimensionality that is certifiable with this method is the maximal $r$ such that
\begin{equation}
\label{Eq2}
r <  \widetilde{F}(\rho,\Psi) \, d+1.
\end{equation}
In our experiment, this allows us to certify the presence of $28$ dimensions of entanglement without the medium. In the presence of the scattering medium and no SLM correction, we obtain a negative lower bound value of the fidelity from the correlation matrices measured in Figures~\ref{Figure3}.c and f, and no entanglement can be certified. After applying the SLM correction, strong correlations in position and momentum are again measurable  (Figs.~\ref{Figure3}.d.g) leading to a lower bound value of the fidelity of $\widetilde{F}(\rho,\Psi)= 0.37$ that allows us to certify $17$ dimensions of entanglement (see Appendix F). 

\begin{figure}
\includegraphics[width=1 \columnwidth]{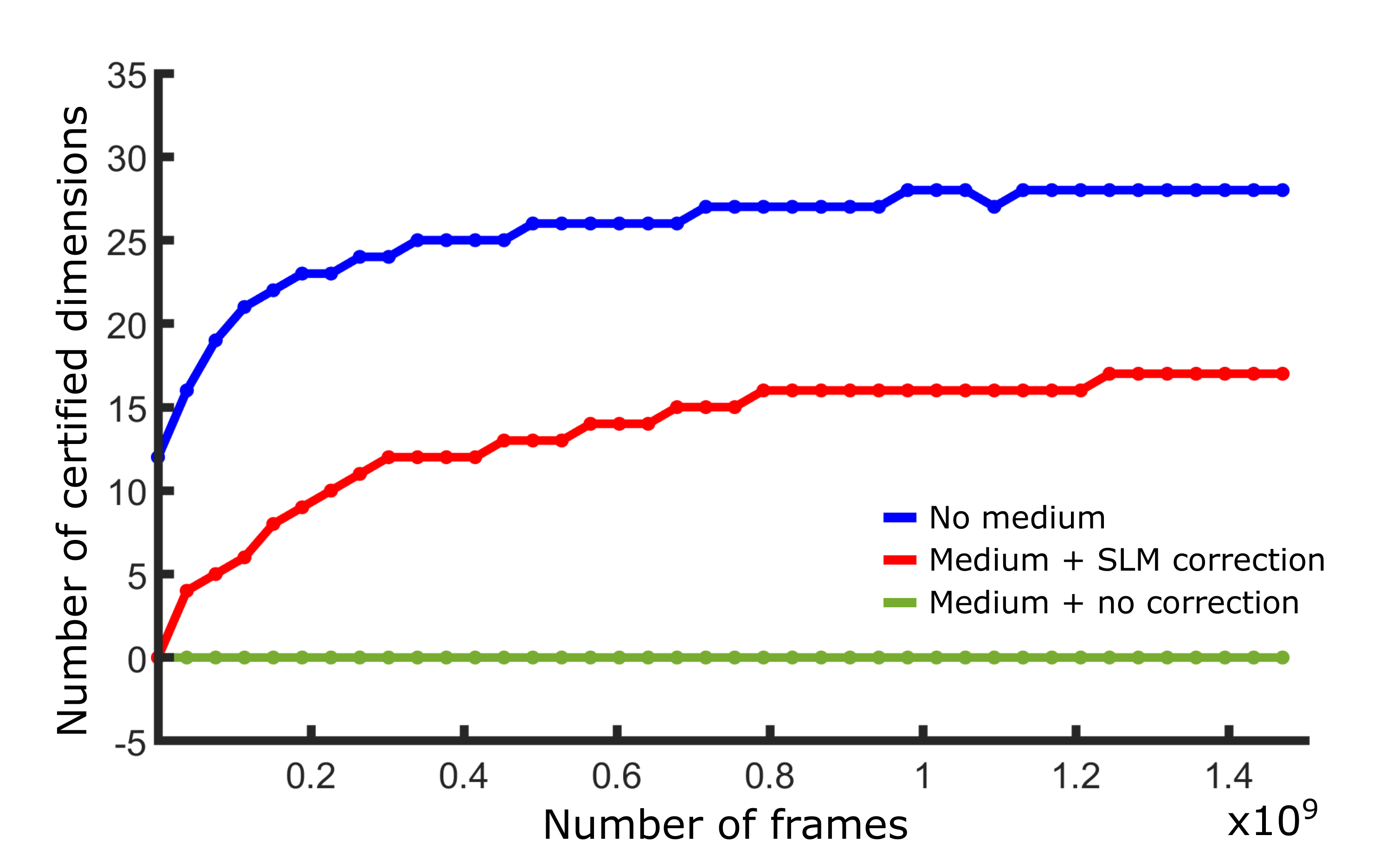} % this command will be ignored
\caption{\label{Figure4} \textbf{Certified dimension analysis.} Number of certified dimensions for different total number of frames without medium (blue curve), with medium and SLM correction (red curve), with medium and no correction (green curve). }
\end{figure}

In practice, the number of dimensions certified at the output depends on our ability to compensate for scattering, but also on the total acquisition time i.e. the number of frames used to compute the correlation matrices~\cite{ndagano_imaging_2020-1}. In order to isolate the role played by wavefront shaping itself, Figure~\ref{Figure4} analyses the number of certified dimensions as a function of the number of frames. While it always remains zero with a scattering medium and no correction (green curve), it increases until it reaches a plateau around $1.2$ billion frames in the case without the medium (blue curve) and with the medium and SLM correction (red curve). The existence of such plateaus is predicted in our simulations (see Appendix H). Because they do not depend on the acquisition time, the values of these plateaus are therefore the relevant quantities to compare in order to evaluate the performance of wavefront shaping for transporting entanglement. In our experiment, we quantify a $39\%$ loss in the number of dimensions. Such a decrease is due to optical losses (i.e. modes not collected at the output) and to the inability for a single SLM to perfectly correct a scattering medium like parafilm which, although relatively thin, is not a simple random phase mask. \\

{\section{Discussion and conclusions}}

In summary, we have shown the transport of high-dimensional spatial entanglement through a scattering medium using a transmission-matrix-based wavefront shaping technique. Using a multi-outcome spatial measurement approach with a SPAD camera, we demonstrated EPR criterion violation with a confidence of $988$ sigma and certified $17$ dimensions of entanglement at the ouptut of the medium. Although our demonstration uses a rather thin scattering medium, the matrix approach (Eq.~\eqref{Eq0}) we introduced can be used with any linear scattering medium. In practice, however, the use of thicker scattering media would spread the correlations over many more spatial modes, many of which would not be collected, making the experiment very challenging~\cite{soro_quantum_2021-1}. Furthermore, finding a solution that satisfies both conditions~\eqref{C1} and~\eqref{C2} to certify the entanglement at the output is not obvious in the general case, especially when using a single SLM to shape light. The use of multi-plane light conversion technologies~\cite{morizur_programmable_2010,fontaine_laguerre-gaussian_2019,lib_reconfigurable_2021,brandt_high-dimensional_2020}, together with more sensitive and faster single-photon cameras such as the next generation SPADs~\cite{madonini_single_2021} or fast time-stamping cameras~\cite{nomerotski_imaging_2019}, could be promising solutions to advance these issues. Simulations provided in the supplementary document section 3 confirms the potential of multi-plane light conversion. Finally, one advantage of our approach is the use of an intense classical light beam to characterize and correct scattering, before switching to the quantum source. This approach could enable similar approaches also in dynamic media, such as biological tissues and atmospheric turbulent layers, which brings us closer to real-world applications. \\

{\section*{Aknowledgements}}

D.F. acknowledges support from the Royal Academy of Engineering Chairs in Emerging Technologies Scheme and funding from the UK Engineering and Physical Sciences Research Council (grants EP/M01326X/1 and EP/R030081/1) and from the European Union's Horizon 2020 research and innovation programme under grant agreement No 801060. S.G. acknowledges funding from the European Research Council ERC Consolidator Grant (SMARTIES-724473). H.D. acknowledges funding from the European Research Council ERC Starting Grant (SQIMIC-101039375). \\

\section*{Authors contributions.} BC performed the experiment. BC, PC and HD analyzed the results. HD conceived the original ideal, designed the experiment and supervised the project. All authors discussed the data and contributed to the manuscript. 

{\section*{Appendix A: Details on the experimental apparatus}} 

The pump is a collimated continuous-wave laser at $405$ nm (Coherent OBIS-LX) with an output power of $100$ mW and a beam diameter of $0.8\pm 0.1$ mm. The SLED is centered at 810nm and has a total bandwdith of approximately 20nm (Superlum). BBO crystal has dimensions $0.5 \times 5 \times 5$ mm and is cut for type I SPDC at $405$ nm with a half opening angle of $3$ degrees (Newlight Photonics). The crystal is slightly rotated around horizontal axis to ensure near-collinear phase matching of photons at the output (i.e. ring collapsed into a disk). A $650$ nm-cut-off long-pass filter is used to block pump photons after the crystals, together with a band-pass filter centered at $810 \pm 5$ nm. The SLM has $1080\times1920$ pixels with a pitch of $8\mu$m and uses a liquid crystal on silicon technology  (Holoeye model Pluto-NIR-II). The SPAD camera is the model SPC3 from Micron Photon Device. It has $32 \times 64$ pixel, a pixel pitch of $45 \mu$m and is operated in the free-running mode with an exposure time of $1 \mu$s. The $4f$ imaging system $f_1-f_2$ in Figure~\ref{Figure1}.a is represented by two lenses for clarity, but is in reality composed of a series of $4$ lenses with focal lengths $50$ mm - $150$ mm - $100$ mm - $200$ mm. The first and the last lens are positioned at focal distances from the crystal and the SLM, respectively, and the distance between two lenses in a row equals the sum of their focal lengths. Similarly, the second $4f$ imaging system $f_3-f_4$ in Figure~\ref{Figure1}.a is composed of a series of $4$ consecutive lenses with focal lengths $200$mm - $100$ mm - $75$ mm - $50$ mm arranged as in the previous case. The other lenses have the following focal lengths: $f_5=30$mm, $f_6=100$mm and $f_7=150$mm. In the momentum basis configuration, the system effective focal length is $75$ mm. In the position basis configuration, the imaging system magnification is $10$.

{\section*{Appendix B: JPD measurement and projections}}

In the experiments shown in our work, the measured JPD $\Gamma$ takes the form of a 4-dimensional matrix containing $(N_Y \times N_X)^4 $ elements, where $N_Y=32$ and $N_X=64$ correspond to the size of the sensor. An element of the matrix is written $\Gamma_{ijkl}$, where $(i,j)$ and $(k,l)$ are pixel labels corresponding to spatial positions $(x_i,y_j)$ and $(x_k,y_l)$. It is measured by acquiring a set of $M+1$ frames $\{I^{(l)} \} _{l \in  [\![ 1,M+1]\!]}$ using a fixed exposure time and then processing them using the formula~\cite{defienne_general_2018,ndagano_imaging_2020-1}: 
\begin{equation}
\label{equ00}
\Gamma_{ijkl} = \frac{1}{M} \sum_{l=1}^M \left[  I^{(l)}_{ij}I^{(l)}_{kl} - I^{(l)}_{ij} I^{(l+1)}_{kl} \right].
\end{equation}
Because the SPAD camera does not resolve the number of photons, photon coincidences at the same pixel (i.e. coefficients $\Gamma_{ijij}$) cannot be measured, and are therefore set to zero. In addition, note that correlation values between neighbouring pixels suffered from cross-talk and must be corrected (see Appendix C). 

$\Gamma_{ijkl}$ is a discrete version of the continuous JPD $\Gamma(\vec{r_1},\vec{r_2}) = |\psi(\vec{r_1},\vec{r_2})|^2$, where $\psi$ is the spatial two-photon wave-function associated with photon pairs, and $\vec{r_1}=x_1 \vec{e_x} + y_1 \vec{e_y}$ and $\vec{r_2}=x_2 \vec{e_x} + y_2 \vec{e_y}$ are transverse spatial positions ($\vec{e_x}$ and $\vec{e_y}$ are unit vectors along the $x$ and $y$ axes, respectively). Such a continuous formalism is used in many theoretical works describing the propagation of spatially-entangled photon pairs~\cite{fedorov_gaussian_2009,schneeloch_introduction_2016}. In our work, to measure the correlations strength in position and momentum, one must project the JPD along the sum- and minus-coordinate (Fig.~\ref{Figure2}). The sum-coordinate $P^+$ and minus-coordinate $P^-$ projections are defined as:

\begin{enumerate}
\item Using the continuous formalism, the sum-coordinate projection $P^+$ is defined as:
\begin{equation}
\label{cont1}
P^+(\vec{r^+}) = \int \Gamma(\vec{r},\vec{r^+}-\vec{r}) d\vec{r}.
\end{equation}
where $\vec{r}= x \vec{e_x}+y \vec{e_y}$. It represents the probability of detecting pairs of photons generated in all symmetric directions relative to the position $\vec{r^+}= x^+ \vec{e_x}+y^+ \vec{e_y}$. In practice, it is calculated using the discrete space formula:
\begin{equation}
\label{equ3s}
P^+_{i^+j^+} = \sum_{i=1}^{N_X} \sum_{j=1}^{N_Y} \Gamma_{(i^+-i) \,(j^+-j) \, i\, j}.
\end{equation}
\item Using the continuous formalism, the minus-coordinate projection $P^-$ is defined as:
\begin{equation}
\label{cont2}
P^-(\vec{r^-}) = \int \Gamma(\vec{r},\vec{r^-}+\vec{r}) d\vec{r}.
\end{equation}
where $\vec{r}= x \vec{e_x}+y \vec{e_y}$. This represents the probability for two photons of a pair to be detected in coincidence between pairs of pixels separated by an oriented distance $\vec{r^-}= x^- \vec{e_x}+y^- \vec{e_y}$. In practice, it is calculated using the discrete space formula:
\begin{equation}
\label{equ3s2}
P^-_{i^-j^-} = \sum_{i=1}^{N_X} \sum_{j=1}^{N_Y} \Gamma_{(i^- +i) \,(j^- +j) \, i\, j}.
\end{equation}
\end{enumerate}

{\section*{Appendix C: Cross-talk correction and hot pixels identification}}

SPAD cameras architecture design involuntary leads to undesirable detection event when a photon or a dark count trigger a pixel~\cite{rech_optical_2008}. As a detection event is created by a charge avalanche due to the first triggering, an electron can reach a neighbouring pixel and trigger it, leading to a cross-talk event. In our work, the average cross-talk probability distribution is first characterized by performing long measurements with the shutter closed (i.e. only dark counts, evenly distributed over the sensor, trigger the pixels), and then removed from the collected data. Furthermore, our SPAD detector exhibits some defective pixels with a higher than normal detection rate~\cite{connolly_hot_2019}. As a result, they lead to multiple fake detection events. In our work, we first identify these hot pixels by performing a long measurement with no light on the sensor and applying a threshold to the resulting image. Values of these pixels are then set to zero in all subsequent acquisitions. More details about cross-talk and hot pixels removal processes are provided in the following.\\

\subsection*{Crosstalk correction process}

\begin{figure}[h]
\centering
\includegraphics[width=0.6 \columnwidth]{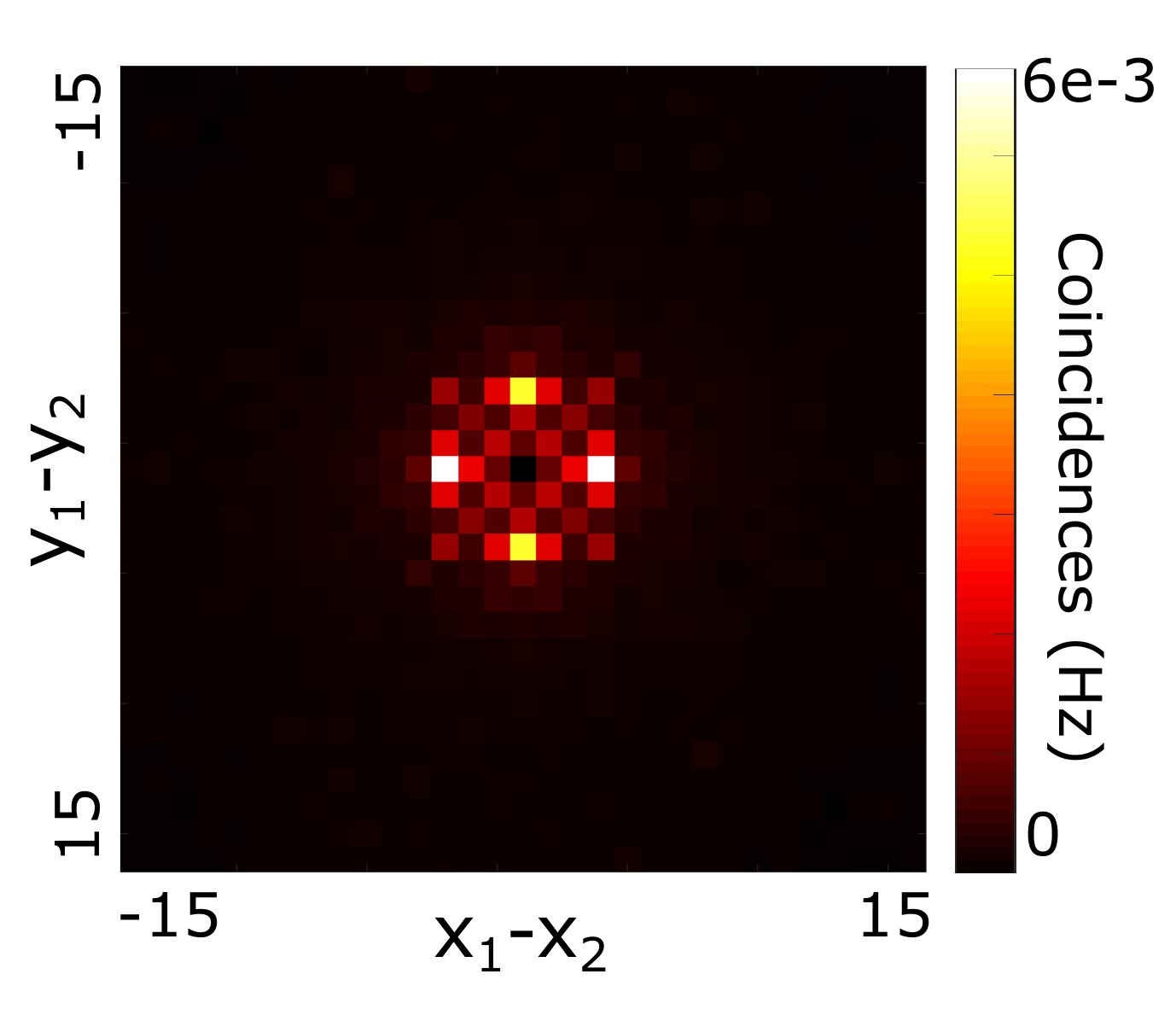} % this command will be ignored
\caption{\label{FigureSM1} \textbf{Cross-talk intensity pattern.} JPD minus-coordinate projection measured with no light falling on the sensor. $7\times10^8$ frames were acquired.}
\end{figure}

Our SPAD camera has some cross-talk issue due to the design of the sensor. As a result, each term of the JPD $\Gamma_{ijkl}$ can be decomposed into two contributions: 
 \begin{equation}
     \Gamma^{(raw)}_{ijkl}=\Gamma^{(opt)}_{ijkl}+\Gamma^{(ct)}_{ijkl},
 \end{equation}
where $\Gamma^{(opt)}_{ijkl}$ is the contribution originating from photon coincidence between pixels $(i,j)$ and $(k,l)$, and $\Gamma^{(ct)}_{ijkl}$ is the contribution originating from cross-talk events between the same pixels. Figure~\ref{FigureSM1} shows the intensity of the cross-talk between two pixels of the sensor in function of the distance between them. It was obtained by measuring the JPD with no light on the sensor (noted $\Gamma^{0}_{ijkl}$) and projecting it along the minus-coordinate. As expected, the cross-talk is non-zero only between pixel pairs that are close to each other, mostly between pixels less than $3$ pixels apart (in both directions).

\begin{figure}[h]
\centering
\includegraphics[width=0.9 \columnwidth]{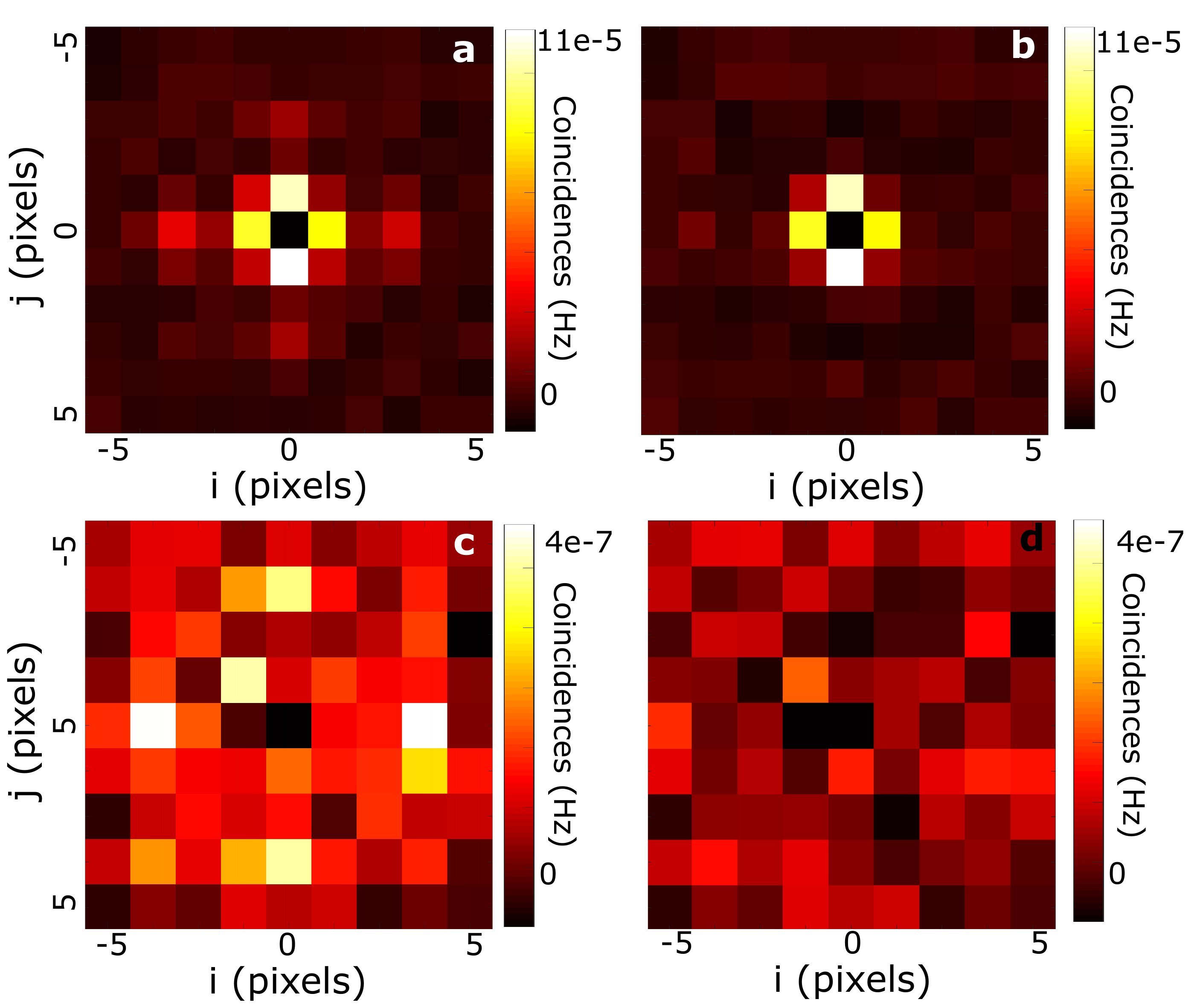} % this command will be ignored
\caption{\label{FigureSM2} \textbf{Conditional probability images before and after cross-talk removal.} \textbf{(a)} $\Gamma^{(raw)}_{ijkl}$ and \textbf{(b)} $\Gamma_{ijkl}$ acquired in the position basis configuration without medium. The reference pixel $(k,l)$ is the central pixel $(0,0)$. \textbf{(c)} $\Gamma^{(raw)}_{ijkl}$ and \textbf{(d)} $\Gamma_{ijkl}$ acquired in the position momentum basis configuration without medium. The reference pixel $(k,l)$ is the central pixel $(0,0)$. $7\times10^8$ frames were acquired.}
\end{figure}

In addition to having a specific shape, cross-talk correlations also depend on the intensity detected on each of the pixels concerned. In short, the more photons are detected by a pair of pixels, the more cross-talk coincidence events are produced. As a result, if the intensity is non-uniform, the cross-talk is spatially dependent. To remove the cross-talk correlations from the measured JPD, we use a correction model that takes into account the intensity dependence : 
\begin{equation}
   \Gamma_{ijkl}=\Gamma^{(raw)}_{ijkl}-\Gamma^{0}_{ijkl} \alpha_{kl} \sqrt{I_{ij}}
    \label{Crosstalk_remove}    
\end{equation}
where $I_{kl}$ is the intensity at pixel $(k,l)$ and $\alpha_{kl}$ is a correction parameter. In a previous work~\cite{eckmann_characterization_2020-2}, cross-talk correction was achieved by setting the parameters $\alpha_{kl} = \alpha \sqrt{I_{kl}}$, where $\alpha$ is a constant. However, the SPAD camera was a different model than ours. In our case, we observed that $\alpha$ could not be set to a constant to correct properly the cross-talk over all the pixels. The reason is probably that the cross-talk pattern shown in Figure~\ref{FigureSM1} is in reality non-uniform across the sensor. To take into account this effect, the parameters $\alpha_{kl}$ are set using the formula:
\begin{equation}
   \alpha_{kl} = \frac{\Gamma^{(raw)}_{ijk(l-3)}+\Gamma^{(raw)}_{ijk(l+3)}}{\Gamma^{0}_{ijk(l-3)}+\Gamma^{0}_{ijk(l+3)}}.
    \label{Crosstalk_remove_2}    
\end{equation}
Here we use the fact that the photon pairs correlation width in the camera plane is much smaller than $135\mu$m$=3$pixels. This means in practice that correlation values between pixels separated by $\pm3$ pixels are only due to cross-talk i.e. $\Gamma^{(raw)}_{ijk(l\pm3)} = \Gamma^{(ct)}_{ijk(l\pm3)}$. In equation~\eqref{Crosstalk_remove_2}, we use these values as reference values to adapt the weight of the correction term $\Gamma^{0}_{ijkl} \sqrt{I_{ij}}$, which enable to compensate for the non-uniformity of the cross-talk shape. Figures~\ref{FigureSM2}.a-d show conditional probability images before ($\Gamma^{(raw)}_{ijkl}$) and after ($\Gamma_{ijkl}$) cross-talk correction. 
   
\subsection*{Hot pixels identification process}

\begin{figure}[h]
\centering
\includegraphics[width=0.9 \columnwidth]{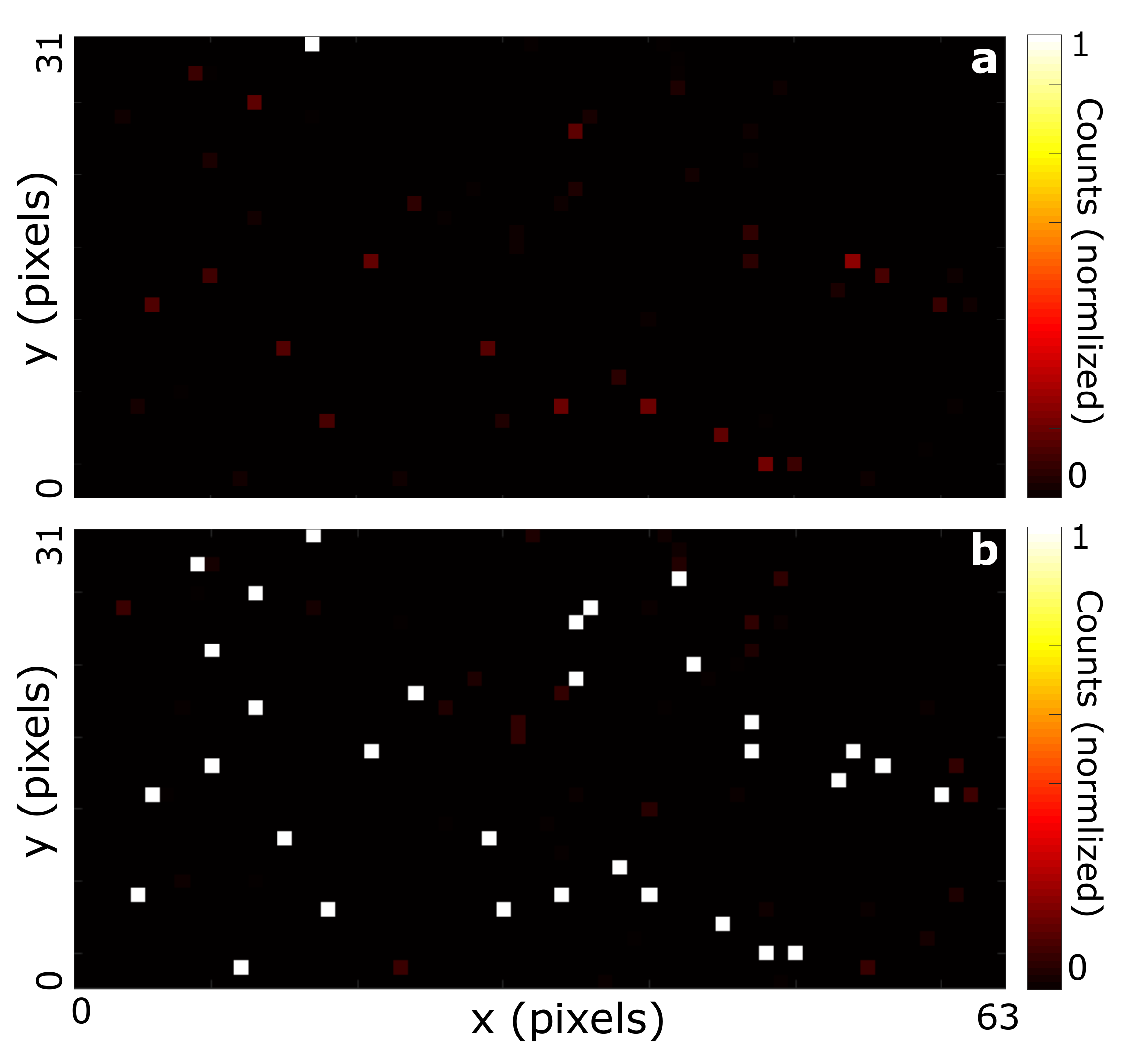} % this command will be ignored
\caption{\label{FigureSM3}\textbf{Hot pixels identification.} \textbf{(a)} Intensity image obtained by summing $1.2\times10^9$ frames acquired with no light on the sensor. \textbf{(b)} Same image with the $32$ identified hot pixels appearing in white.}
\end{figure}

To remove the hot pixels, we first acquired $1.2\times10^9$ frames with no light on the sensor and sum them to obtain the intensity image shown in Figure~\ref{FigureSM3}. We then apply a threshold to the image: all pixels with value above $10\%$ of the maximum are identified as hot pixels and set to zero in all further acquisitions. In total, $32$ pixels over a total of $2048$ were identified.
 
\section*{Appendix D: Gaussian model of the two-photon wavefunction} 

The two-photon wave-function $\psi(\vec{r_1},\vec{r_2})$ associated with photon pairs produced by type-I SPDC at the surface of a thin non-linear crystal can be approximated using a Gaussian model~\cite{schneeloch_introduction_2016,fedorov_gaussian_2009}:
\begin{equation}
\label{gaussian}
\psi(\vec{r_1},\vec{r_2}) = A \, \exp{ \left( \frac{-|\vec{r_1}-\vec{r_2}|^2}{4 \sigma_{\vec{r}}^2} \right)} \exp{\left( \frac{-|\vec{r_1}+\vec{r_2}|^2 \sigma_{\vec{k}}^2 }{4}\right)}
\end{equation}
where $\sigma_{\vec{r}}$ and $\sigma_{\vec{k}}$ are the position and momentum correlation widths of photons at the output of the crystal, respectively, and $A$ is a constant. As a result, minus- and sum-coordinate JPD projections measured in the position and momentum basis, respectively, also have a Gaussian shapes i.e. $\exp(-|\vec{r_1-r_2}|^2/2\sigma_\vec{r}^2)$ and $\exp(-|\vec{k_1}+\vec{k_2}|^2/2\sigma_\vec{k}^2)$. Under this approximation, the correlation widths $\sigma_\vec{r}$ and $\sigma_\vec{k}$ are exactly identified with the uncertainties $\Delta \vec{r} = \Delta (\vec{r_1}-\vec{r_2})$ and $\Delta \vec{k} = \Delta (\vec{k_1}+\vec{k_2})$~\cite{giovannetti_characterizing_2003,howell_realization_2004}. This justifies the use of a Gaussian model to extract the values of $\Delta \vec{r}$ and $\Delta \vec{k}$, as shown in Figure~\ref{Figure2}. However, it should be kept in mind that if the two-photon wavefunction can no longer be modeled by a Gaussian, as is the case with the medium and no SLM correction (Fig.~\ref{Figure2}.d), then the correspondence with $\Delta \vec{r}$ and $\Delta \vec{k}$ values is no longer exact. This issue would also occur in the case of thicker scattering media, as we expect the SLM to refocus coincidences in the central peak but with a non-zero coincidence background surrounding it.

\section*{Appendix E: Confidence level on EPR criterion violation}

Equation~\eqref{Eq0} shows the EPR separability criterion considered in our work~\cite{giovannetti_characterizing_2003}. This criterion was used in previous works to demonstrate the presence of entanglement between photons with different types of detectors, including scanning single-pixel detectors~\cite{howell_realization_2004}, s-cMOS camera~\cite{dabrowski_einsteinpodolskyrosen_2017}, electron-multiplying CCD (EMCCD) camera~\cite{moreau_realization_2012,edgar_imaging_2012} and SPAD camera~\cite{ndagano_imaging_2020-1}. To achieve EPR criterion violation, transverse position and momentum correlation widths, $\Delta\vec{r}$ and $\Delta\vec{k}$, are first estimated by fitting the sum- and minus-coordinate projections of the momentum and position bases configurations by a Gaussian model~\cite{fedorov_gaussian_2009,moreau_realization_2012,edgar_imaging_2012} of the form $f(r)=$ $a\exp(-r^2/2\Delta^2)$, where $a$ is a fitting parameter and $\Delta$ is the desired correlation width value $\Delta\vec{r}$ or $\Delta\vec{k}$ (see Appendix D for more details about the Gaussian model). The presence of noise in the sum- and minus-coordinate images induce uncertainties on values $\Delta\vec{r}$ and $\Delta\vec{k}$ returned by the fitting process. The standard deviation of the noise $\Sigma$ is measured in an area composed of $15 \times 15$ pixels surrounding the central peak of coincidence. The link between the correlation width uncertainty $\delta_\Delta$ ($\delta_\Delta = \delta_{\Delta\vec{r}}$ or $\delta_{\Delta\vec{k}}$) and $\Sigma$ is given by calculating the value of $grad[f]$ a the position $r = \Delta$: $\left| \frac{df}{dr}(r=\Delta) \right| = a/(\Delta \sqrt{e}).$ Then, expanding it at the first order in $r$: $\delta f = a \delta r /(\Delta \sqrt{e}).$ In our case, the variations $\delta f$ and $\delta r$ identify to the uncertainty quantities $\Sigma$ and $\delta_\Delta$, respectively, which finally leads to:
	\begin{equation}
	\delta_\Delta = \frac{\Sigma \sqrt{e}\Delta}{a}.
	\label{deltaDelta}
	\end{equation}
	All correlation width values and uncertainties are expressed in the coordinate system of the crystal, after taking into consideration the magnifications introduced by the imaging systems detailed in Figure~\ref{Figure1}.a. Then, the confidence level of the EPR violation is defined as:
	\begin{equation}
	\label{equconfidence}
	C=\frac{|1/2-\Delta\vec{r}\cdot \Delta\vec{k} |}{\sigma}
	\end{equation}
	where $\sigma = \Delta\vec{r}\cdot \Delta\vec{k} \sqrt{(\delta_{\Delta \vec{k}} / \Delta \vec{k})^2+(\delta_{\Delta \vec{r}} / \Delta \vec{r})^2} $ is the uncertainty on the product $\Delta\vec{r}\cdot \Delta\vec{k}$. The confidence level essentially expresses the deviation of the measured value $\Delta \textbf{r} \Delta \textbf{k}$ from the violation limit in number of $\sigma$.

\section*{Appendix F: Entanglement certification}

In our experiment, we use discrete transverse position and momentum bases given by a set pixels defined on the SPAD camera and noted $\{ \ket{m}\}_{m \in [\![1;d]\!]}$ and $ \{ \ket{\tilde{p}} \} _{p \in [\![1;d]\!]} $, respectively. Our approach is based on the protocol proposed by Erker \textit{et al.}~\cite{erker_quantifying_2017} where these bases are used as two mutually unbiased bases (MUBs) to certify high-dimensional entanglement. They are linked according to:
	\begin{equation}
	\ket{\tilde{p}} = \frac{1}{\sqrt{d}} \sum_{m=0}^{d-1} \omega^{km} \ket{m} 
	\end{equation} 
	where $\omega = e^{2 \pi i / d}$. Experimentally, these bases are accessed using lenses to image or Fourier-image the output of the non-linear crystal. A subset of pixels is then selected in the illuminated areas of the sensor to optimize coincidence signals measured. In our case, we selected 45 pixels evenly separated from each other by $2$ pixels and located on a disk. One difference between the scheme of Eker \textit{et al.}~\cite{erker_quantifying_2017} and our work is that only one image is produced on the camera in our case, against two in their proposal using a beam splitter. Our setup prevents us from accessing the coincidence rate at the same pixel i.e. same spatial modes. Instead, the coincidence rate of photon pairs in the same pixel is inferred by measuring the coincidence rate between each pixel and its neighbour. This inference leads to a lower value of the intra-pixel coincidence rate in the position basis, which therefore underestimates the real entanglement witness value.\\
To certify the presence of high-dimensional entanglement in the measured state $\rho$, we employ a recently developed witness that uses correlations in at two MUBs~\cite{bavaresco_measurements_2018} i.e. $\{ \ket{m}\}_{m \in [\![1;d]\!]}$ and $ \{ \ket{\tilde{p}} \} _{p \in [\![1;d]\!]}$. Using coincidence measurements in these two bases, one can determine a lower bound for the fidelity $F(\rho,\Phi)$ of the state $\rho$ to a pure bipartite maximally entangled target state $\ket{\Phi}$. Since the fidelity to a target entangled state also provides information about the dimensionality of entanglement, we use this bound for certifying the dimension of entanglement of the state produced in our experiment. We consider a maximally entangled target state written as:
\begin{equation}
\ket{\Phi} = \frac{1}{\sqrt{d}}\sum_{m=0}^{d-1} \ket{mm}
\end{equation}
with $d=45$. The fidelity $F(\rho,\Phi)$ of the state $\rho$ to the target state $\ket{\Phi}$ is defined as:
\begin{eqnarray}
\label{fidelity}
F(\rho,\Phi) &&= \mbox{Tr} \left(\ket{\Phi} \bra{\Phi} \rho \right) \nonumber \\
&&= \sum_{m,n=0}^{d-1} \brakket{mm}{\rho}{nn} \nonumber \\
&& = F_1(\rho,\Phi)+F_2(\rho,\Phi)
\end{eqnarray}

where 
\begin{eqnarray}
 F_1(\rho,\Phi) &=& \sum_{m=0}^{d-1} \brakket{mm}{\rho}{mm} \label{F1} \\
 F_2(\rho,\Phi) &=& \sum_{m\neq n}^{d-1} \brakket{mm}{\rho}{nn} \label{F2}
\end{eqnarray}

The entanglement dimensionality can be deduced from the fidelity taking into account that for any state $\rho$ of Schmidt number $r \leq d$, the fidelity of Eq.~\eqref{fidelity} is bound by:
\begin{equation}
\label{fidelity2}
F(\rho,\Phi) \leq B_r(\Phi) = \frac{r}{d} 
\end{equation}
Hence, any state with $F(\rho,\Phi) > B_r(\Phi)$ must have an entanglement dimensionality of at least $r+1$. Our goal is therefore to obtain a lower bound on the fidelity as large as possible for the target state whose Schmidt rank is as close as possible to the local dimension $d$. To achieve this experimentally, the method described in~\cite{bavaresco_measurements_2018} works the following way:\\
\noindent Step 1: Matrix elements $ \{ \brakket{mn}{\rho}{mn} \}_{m,n}$ are calculated from the coincidence counts $\{N_{mn}\}_{mn}$ measured in the discrete position basis via:
\begin{equation}
\brakket{mn}{\rho}{mn} = \frac{N_{mn}}{\sum_{k,l}N_{kl}} 
\end{equation}
These elements are shown in the matrix in Figure~\ref{Figure3}.a.b.c They enable to calculate directly the term $F_1(\rho,\Phi)$ from the definition given by Eq~\eqref{F1}.\\
\noindent Step 2: Matrix elements $ \{ \brakket{\tilde{p} \tilde{v}}{\rho}{\tilde{p} \tilde{v}} \}_{pv}$ are calculated from the coincidence counts $\{\tilde{N}_{pv}\}_{pv}$ measured in the discrete momentum basis via:
\begin{equation}
\brakket{\tilde{p} \tilde{v}}{\rho}{\tilde{p} \tilde{v}} = \frac{\tilde{N}_{pv}}{\sum_{k,l}\tilde{N}_{kl}} 
\end{equation}
These elements are shown in the matrices in Figure~\ref{Figure3}.d.e.f These matrix elements, together with those of the discrete position basis, allow us to bound the fidelity term $F_2(\rho,\Phi)$. This lower bound $\tilde{F}_2(\rho,\Phi)$ is calculated via:
\begin{eqnarray}
&&\tilde{F}_2(\rho,\Phi) =  \sum_{p=0}^{d-1} \brakket{\tilde{p} \tilde{p}}{\rho}{\tilde{p} \tilde{p}} - \frac{1}{d} - \nonumber \\
&& \sum_{\substack{m \neq n',m \neq n  \\ n \neq n',n' \neq m' }} \gamma_{mnm'n'} \sqrt{\brakket{mn}{\rho}{mn} \brakket{m'n'}{\rho}{m'n'} } \label{f2tilde}
\end{eqnarray}
where the prefactor $\gamma_{mnm'n'}$ is given by
\begin{equation}
\gamma_{mnm'n'} = \left\{
    \begin{array}{ll}
        0 & \mbox{if } (m-m'-n+n') \mbox{ mod } d\neq0 \\
        \frac{1}{d} & \mbox{otherwise.}
    \end{array}
\right.
\end{equation}
A derivation of Eq.~\eqref{f2tilde} can be found in the Methods section of~\cite{bavaresco_measurements_2018}.\\
\noindent Step 3: A lower bound on entanglement is calculated as $\tilde{F}(\rho,\Phi) = F_1(\rho,\Phi)+\tilde{F}_2(\rho,\Phi) \leq F(\rho,\Phi)$. This lower bound value is finally compared to the certification bound $B_r(\Phi)$ as
\begin{equation}
B_r(\Phi) < \tilde{F}(\rho,\Phi) \leq B_{r+1}(\Phi)
\end{equation}
thus certifying entanglement in $r+1$ dimensions.\\
Using this approach, we note that no assumptions are directly made about the underlying quantum state $\rho$. However, assumptions are made about our measurement process. Indeed, by using Eq.~\eqref{equ00} to measure the JPD of photon pairs, we effectively perform a subtraction of accidental counts. Correcting for accidental coincidence is acceptable in our experiment since we trust our measurement devices and the final goal is only to assess the presence of entanglement and its dimension. However, such an assumption would not be acceptable in an adversarial scenario such as quantum key distribution as it is likely to compromise the security of the protocol. 

\begin{figure}[h]
\centering
\includegraphics[width=1 \columnwidth]{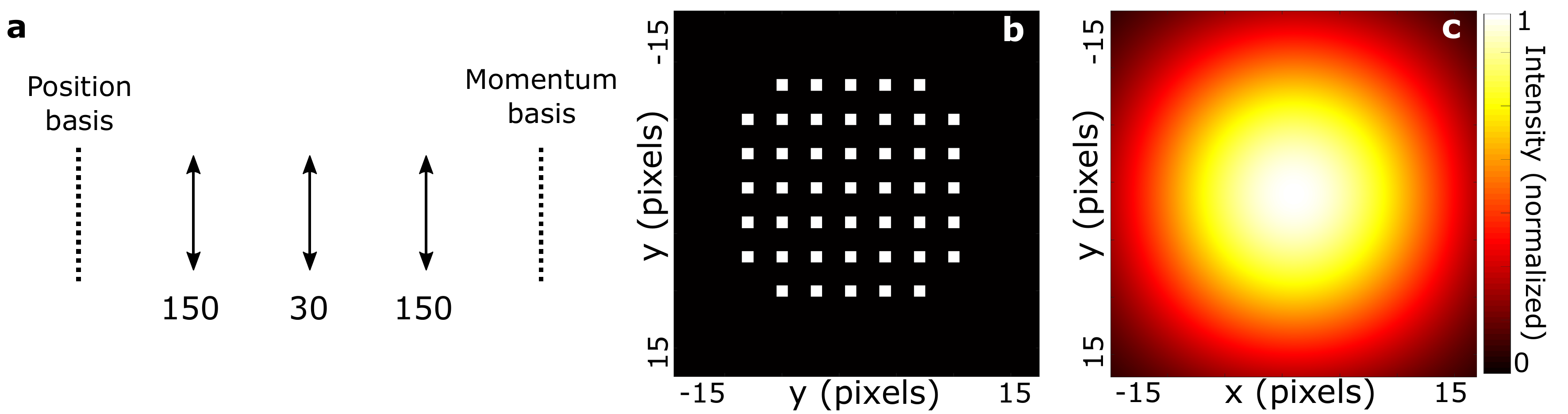} % this command will be ignored
\caption{\textbf{Measure of unbiasedness.}\textbf{a.} Experimental configuration representing the optical transformation that links the position and momentum bases in our experiment. \textbf{b.} Spatial arrangement of the $45$ selected pixels in the position and momentum optical planes. \textbf{c.} Sinc-shaped intensity distribution produced by a pixel located in one optical plane into the reciprocal optical plane.
\label{FigureSM6}}
\end{figure}

In addition,  the sets of pixels selected to form the discrete position and momentum bases (Figs~\ref{Figure3}.a.b) are not perfectly mutually unbiased. Indeed, one set is obtained from the other by applying a continuous Fourier transform to a discrete set of modes, which is formaly different than using a discrete Fourier transform. In our experimental configuration, however, the deviation from perfect unbiasedness is very small. To quantify it, we use an unbiasedness quantifier $E_n$ defined as~\cite{tasca_mutual_2018}:
\begin{equation}
E_n = \sum_{\tilde{v}=0}^{d-1} p_{\tilde{v}|n} \log_2(p_{\tilde{v}|n}).
\end{equation}
where $p_{\tilde{v}|n} = p_{\tilde{v}n} / \sum_{\tilde{v}} p_{\tilde{v} n}$, with $p_{\tilde{v}n} = |\langle \tilde{v} | \rho | n \rangle|^2$, $\{ \ket{n}\}_{n \in [\![1;d]\!]}$ and $ \{ \ket{\tilde{v}} \} _{\tilde{v} \in [\![1;d]\!]} $ are the discrete transverse position and momentum bases, and $d=45$ is the dimension of the subspace. To estimate $p_{\tilde{v}n}$, we consider the optical configuration shown in Figure~\ref{FigureSM6}.a. Such a configuration represents the effective optical transformation between the position and momentum bases in our experimental setup (Figure~\ref{Figure1}). In practice, $p_{\tilde{v}n}$ can be seen as the intensity measured at the discrete position $\ket{\tilde{v}}$ in the momentum basis when a coherent source emits light from the position $\ket{{n}}$ in the position basis. In our experiment, values $p_{\tilde{v}n}$ can therefore be estimated from the spatial arrangement of all the pixels forming each basis and their shape. Figure~\ref{FigureSM6}.b shows the spatial arrangement of the pixels in the two optical planes. It is the same in the momentum and position basis optical planes. In addition, each pixel is a square. The intensity produced by each pixel in the other basis has therefore the shape of a bi-dimensional sinc function and is independant of the pixel position. Figure~\ref{FigureSM6}.c shows such an intensity distribution. Superimposing Figure~\ref{FigureSM6}.b and Figure~\ref{FigureSM6}.c enables to estimate the values $p_{\tilde{v}|n}$ for all $n \in [\![1;d]\!]$ and $\tilde{v} \in [\![1;d]\!]$. In our experiment, we then estimate $E_n = E = 5.479$. This value only differs by $0.5\%$ from the maximum value of $\log_2(45) \approx 4.592$. 

\section*{Appendix G: Entangled-photon matrix formalism}

Under the paraxial approximation, propagation of spatially-entangled photon pairs through a linear optical system is described with the formula~\cite{abouraddy_entangled-photon_2002-2}:
\begin{equation}
\label{2pcont}
    \psi^{out}(\vec{r_1'},\vec{r_2'}) = \iint h(\vec{r_1'},\vec{r_1}) h(\vec{r_2'},\vec{r_2}) \psi^{in}(\vec{r_1},\vec{r_2}) d\vec{r_1} d\vec{r_2},
\end{equation}
where $h(\vec{r'},\vec{r})$ is the impulse response function of the linear system, $\psi^{out}(\vec{r_1'},\vec{r_2'})$ and $\psi^{in}(\vec{r_1},\vec{r_2})$ are the spatial two-photon wavefuntions in the output and input planes of the system, respectively, and $\{\vec{r_i}\}_{i=1,2}$ and $\{\vec{r_i}'\}_{i=1,2}$ are the transverse spatial coordinate in the input and output planes, respectively. Using the same approach as for the propagation of coherent light~\cite{popoff_measuring_2010}, equation~\eqref{2pcont} can be discretized into the following matrix form 
\begin{equation}
\label{2pdis}
    \Psi^{out} = H \Psi^{in} H^t,
\end{equation}
where $H$ is the transfer matrix of the optical system i.e. a discrete version of $h$, and $\Psi^{out}$ and $\Psi^{in}$ are matrices that represent discrete forms of the two-photon wavefuntions $\psi^{out}$ and $\psi^{in}$, respectively. Note that, as in the classical matrix approach, the input and output planes are discretized as pixels (i.e. optical modes) which are then linearly ordered. This allows the use of matrices even if the associated functions $h$, $\psi^{out}$ and $\psi^{in}$ are quadrivariate functions. In our experiment, $H=TD$, where $T$ is the transmission matrix of the scattering medium and $D$ the diagonal matrix associated with the SLM. In our work, we also use the approximation $\psi^{in} \approx 1\!\!1$ to obtain conditions~\eqref{C1} and~\eqref{C2}. Indeed, photon pairs at the output surface of the crystal are strongly correlated in positions. We estimated the corresponding position correlation in the crystal plane to $4 \mu$m using an EMCCD camera~\cite{moreau_realization_2012,edgar_imaging_2012}. Accounting for the magnification between the crystal and the SLM, the correlation width in the SLM plane is about $24\mu$m, which is much smaller than the size of a macro-pixel used to measure the transmission matrix ($120 \mu$m). When discretized in the SLM macro-pixel basis, the matrix $\Psi^{in}$ has therefore negligible off-diagonal components compared to its diagonal components, which allows us to approximate it to the identity matrix.

\section*{Appendix H: Simulations}

Three types of simulations were conducted to support our experimental results. First, simulations based on matrix multiplications were used to predict $\Psi^{out}$. Equation~\eqref{Eq0} was computed using an experimentally measured transmission matrix, verifying that the correction SLM pattern (Fig.~\ref{Figure1}.c) allows $\Psi^{out}$ to satisfy conditions~\eqref{C1} and~\eqref{C2}. Second, simulations were performed in the general case of a thick scattering medium. They confirmed the possibility to use our approach in this case and the interest of using multi-plane light conversion techniques. Third, simulations were performed to confirm the existence of a plateau at large number of frames in the curves shown in Figure~\ref{Figure4}.  More details about these simulations are provided in the following:

\subsection*{Simulation of $\Psi^{out}$ using the entangled-photon matrix approach.}

\begin{figure}
\centering
\includegraphics[width=1 \columnwidth]{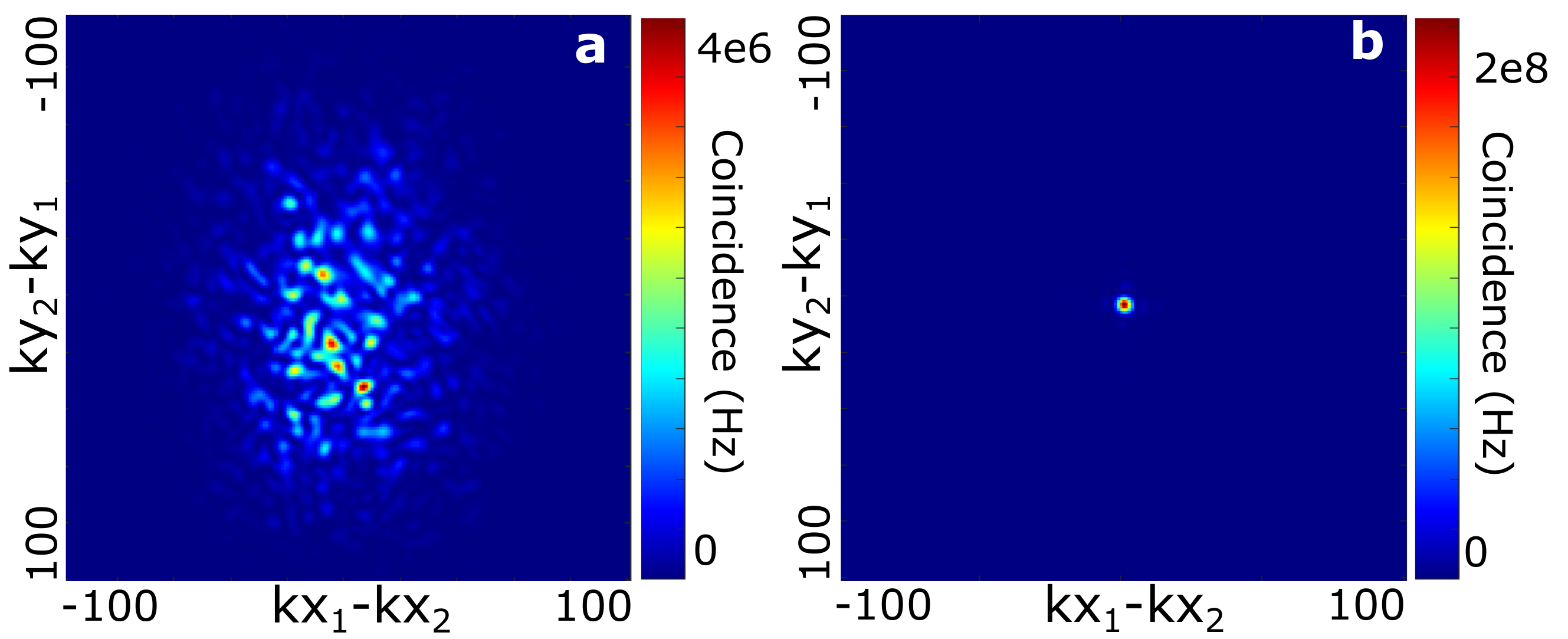} \caption{{\textbf{Simulated two-photon output fields.} Sum-coordinate projections of $|\Psi_{out}|^2$ obtained with \textbf{(a)} a flat phase pattern and the \textbf{(b)} correction phase pattern. Spatial coordinate are in pixels.}
\label{FigureSM4}}
\end{figure}

Figure~\ref{FigureSM4} shows simulation results demonstrating the focusing of spatial correlations in the momentum basis through the scattering medium. The main steps of the algorithm to obtain these results are described in the following: 
\begin{enumerate}
\item \textit{Two-photon input field.} Exploiting that in our experiment the correlation width of photon pairs in the SLM plane ($24\mu$m) is much smaller than the size of a macro-pixel of the SLM ($120\mu$m), the input two-photon wave-function written in the SLM macro-pixel basis is approximated by an identity matrix. In Matlab, it takes the form of a $1024\times1024$ identity matrix noted $\Psi_{in}$. 
\item \textit{Propagation.} The two-photon field after propagation through the medium is obtained using the following equation: 
\begin{equation}
\Psi_{out} = TD \Psi_{in} (TD)^T 
\end{equation} 
where $\Psi_{out}$ is the matrix associated with the two-photon output field, $T$ is the transmission matrix of the system and $D$ is the diagonal matrix associated with the SLM. $T$ was measured between $32\times32=1024$ macro-pixels of the SLM and $100\times100$ pixels of the CCD camera. Since we experimentally measured it using a co-propagating reference, the transmission matrix obtained slightly differs from the real transmission matrix $T$ i.e. it is in reality $D' T$, where $D'$ is a complex diagonal matrix with a unknown speckle pattern on its diagonal~\cite{popoff_measuring_2010}. However, because the SLM acts only at the input of the medium, the presence of $D'$ does not impact our simulations. For clarify, the measured transmission matrix will then be written $T$. %$D$ contains $1024$ complex phase terms $\{ e^{i \theta_k} \}_{k \in [\![ 1,1024 ]\!]}$, where $\theta_k$ is the phase term associated with the $k^{th}$ SLM macro-pixel. 
\end{enumerate}
In this simulation, two different types of images were produced at the output for two different phase patterns programmed on the SLM. Figure~\ref{FigureSM4}.a shows the sum-coordinate projection of $|\Psi_{out}|^2$ obtained with a flat phase pattern i.e. $\theta_k=0$ for all $k\in [\![ 1,1024 ]\!]$. It shows a speckle pattern similar to the one observed experimentally in Figure~\ref{Figure2}.d, albeit with a higher pixel resolution. Figure~\ref{FigureSM4}.b shows the sum-coordinate projection of $|\Psi_{out}|^2$ obtained with the phase pattern $\theta_k = \arg(T_{pk}^*)$, for  $k\in [\![ 1,1024 ]\!]$, with $p$ being the central pixel of the camera. This phase pattern is shown in Figure~\ref{Figure1}.c. In this case, we observe the presence of a strong peak of coincidence in the output sum-coordinate projection, showing that spatial correlations are restored at the output when programming this phase pattern. It corresponds to the experimental results shown in Figure~\ref{Figure2}.f. 

\subsection*{Simulations of entanglement manipulation through a thick scattering medium.}

In this section, we simulate the propagation and manipulation of spatially-entangled photon pairs through a thick scattering medium. In such a general case, equations (2) and (3) cannot be simplified and perfectly satisfied using a single SLM. Nevertheless, using an optimisation approach, one can try to get as close as possible to the ideal situation. In the following example, we describe a computer-based optimization algorithm that re-focuses photons pairs in coincidence in both the position and momentum basis. 

\begin{figure}
\centering
\includegraphics[width=1 \columnwidth]{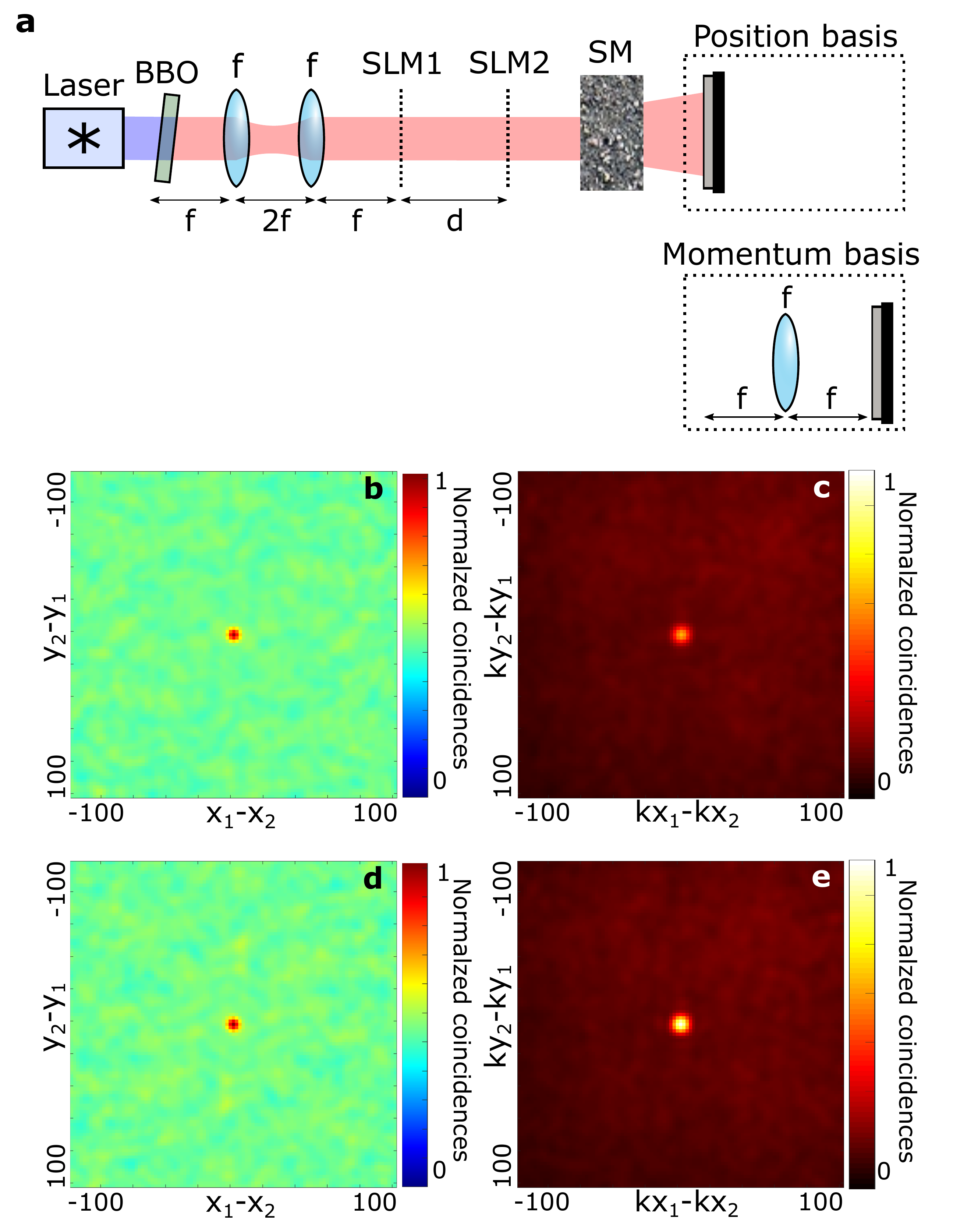}. \caption{\textbf{Simulation of entangled photons propagating through a thick scattering medium.} \textbf{(a)} Experimental setup considered in our simulation. Pump beam has a wavelength of $405$nm and a beam diameter of $5$mm. Focal length is $f=500$mm and propagation distance $d=200$mm. The non-linear crystal is considered infinitely thin. \textbf{(b)} Minus-coordinate projection measured in the position basis configuration and \textbf{(c)} sum-coordinate projection measured in the momentum basis configuration after optimization usin SLM1. \textbf{(d)} Minus-coordinate projection measured in the position basis configuration and \textbf{(e)} sum-coordinate projection measured in the momentum basis configuration after optimization using SLM1 and SLM2. SM: Scattering medium ; SLM: Spatial light modulator; BBO:  $\beta$-barium borate.  
\label{FigureSM7}}
\end{figure}

Figure~\ref{FigureSM7}.a shows the experimental setup associated with our simulation. Entangled photons propagating in the position and momentum basis configurations are described by the two following equations:
\begin{eqnarray}
T D_2 P_d D_1^2 P_d^t D_2 T^t &=& \Psi^{out}_{p}, \\
F T D_2 P_d D_1^2 P_d^t D_2 T^t F^t &=& \Psi^{out}_{m}. 
\end{eqnarray} 
$D_1$ and $D_2$ are diagonal matrices associated with SLM1 and SLM2, respectively, $P_d$ is the matrix associated with free-space propagation over the distance d, $\Psi^{out}_{p}$ and $\Psi^{out}_{m}$ are the two two-photon output fields in the position and momentum basis configurations, respectively, and $T$ is the transmission matrix of the scattering medium. $T$ can be generated numerically as in i.i.d complex gaussian matrix, as we did to obtain the results shown in Figure~\ref{FigureSM7}.b-e, or it could alternatively be measured experimentaly using classical light (as in~\cite{popoff_measuring_2010}) and then loaded onto the computer. To perfom the optimization approach, we choose as an optimization target the sum of the coincidence rates taken at the center of the minus-coordinate projection in the position basis and this at the center of the sum-coordinate projection in the momentum basis i.e. $P_{p,00}^- + P_{m,00}^+$. $P_{p,00}^-$ and $P_{m,00}^+$ are defined using equations~\ref{equ3s} and~\ref{equ3s2} as follow:
\begin{eqnarray}
P_{p,00}^- &=& = \sum_{i=1}^{N_X} \sum_{j=1}^{N_Y} |\Psi^{out}_{m}|^2_{i \,j \, i\, j}, \\
P_{m,00}^+ &=& = \sum_{i=1}^{N_X} \sum_{j=1}^{N_Y} |\Psi^{out}_{p}|^2_{-i \,-j \, i\, j}, 
\end{eqnarray}
where $N_X$ and $N_Y$ are the number of elements in each spatial axis used in our simulation. 
When performing the optimisation using only SLM1 (i.e. SLM2 remains flat), we observe the apparition of a peak of coincidences in both the minus-coordinate projection in the position basis (Fig.~\ref{FigureSM7}.b) and the sum-coordinate projection in the momentum basis (Fig.~\ref{FigureSM7}.c). These results show that our optimization approach enables to re-focus entangled photons in coincidences in both bases simultaneously. To improve the refocusing process, we also inserted a second SLM (SLM2) in the system at a distance $d$ from SLM1. After optimisation, Figures~\ref{FigureSM7}.d and e show an improvement in the peak-to-background ratio in both projections. 

\begin{figure}[h]
\centering
\includegraphics[width=1 \columnwidth]{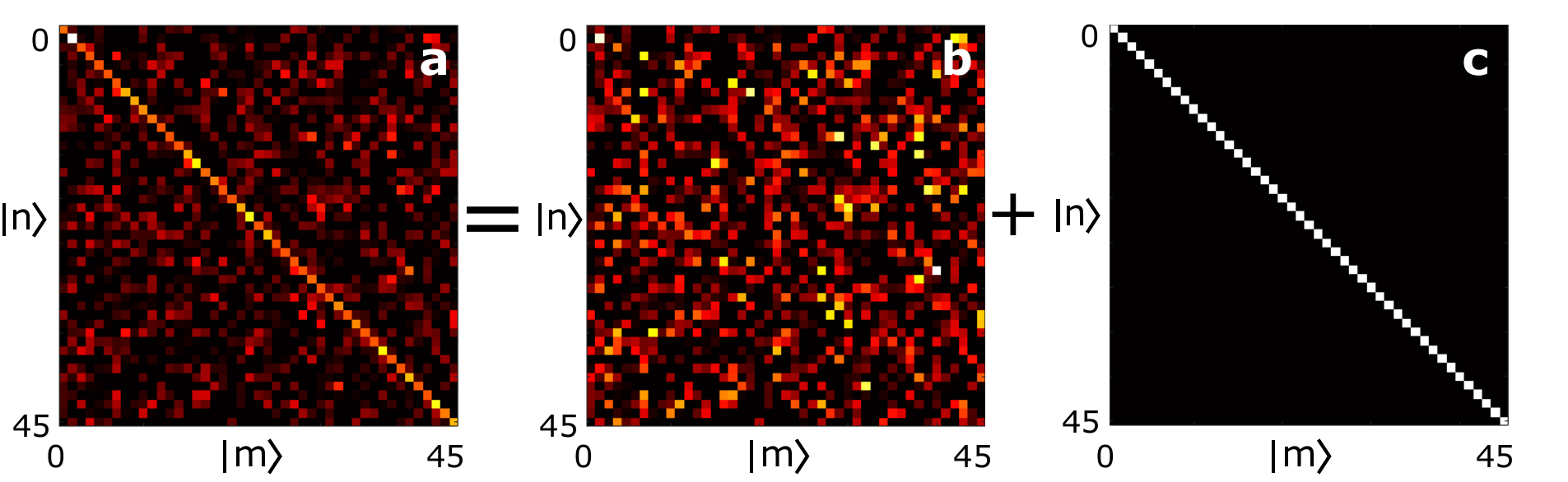} % this command will be ignored
\caption{\textbf{Simulation of a correlation matrix.} In the case of the maximally-entangled state, a \textbf{(a)} correlation matrix is obtained by adding a \textbf{(c)} $45 \times 45$ matrix made of random values with mean parameter $0$ and standard deviation $\sigma = 1/\sqrt{N}$, where $N$ is the number of frames, to a \textbf{(c)} $45 \times 45$ diagonal matrix with diagonal values $\alpha$. In this example, $N=10^6$ and $\alpha = 1/45$.
\label{FigureSM4bis}}
\end{figure}

These simulation results confirm that our wavefront shaping approach can also be used to manipulate entanglement through thick scattering media. In particular, the proposed algorithm enalbles to optimize coincidences in two different output bases simultaneously, an essential step towards high-dimensional entanglement certification. In addition, we also show that using multiple SLMs (i.e. multi-plane light conversion) is a promising method to achieve such a task.

\subsection*{Simulation for studying the variation of the fidelity and number of certified dimensions with the number of frames.}

\begin{figure}[h]
\centering
\includegraphics[width=0.8 \columnwidth]{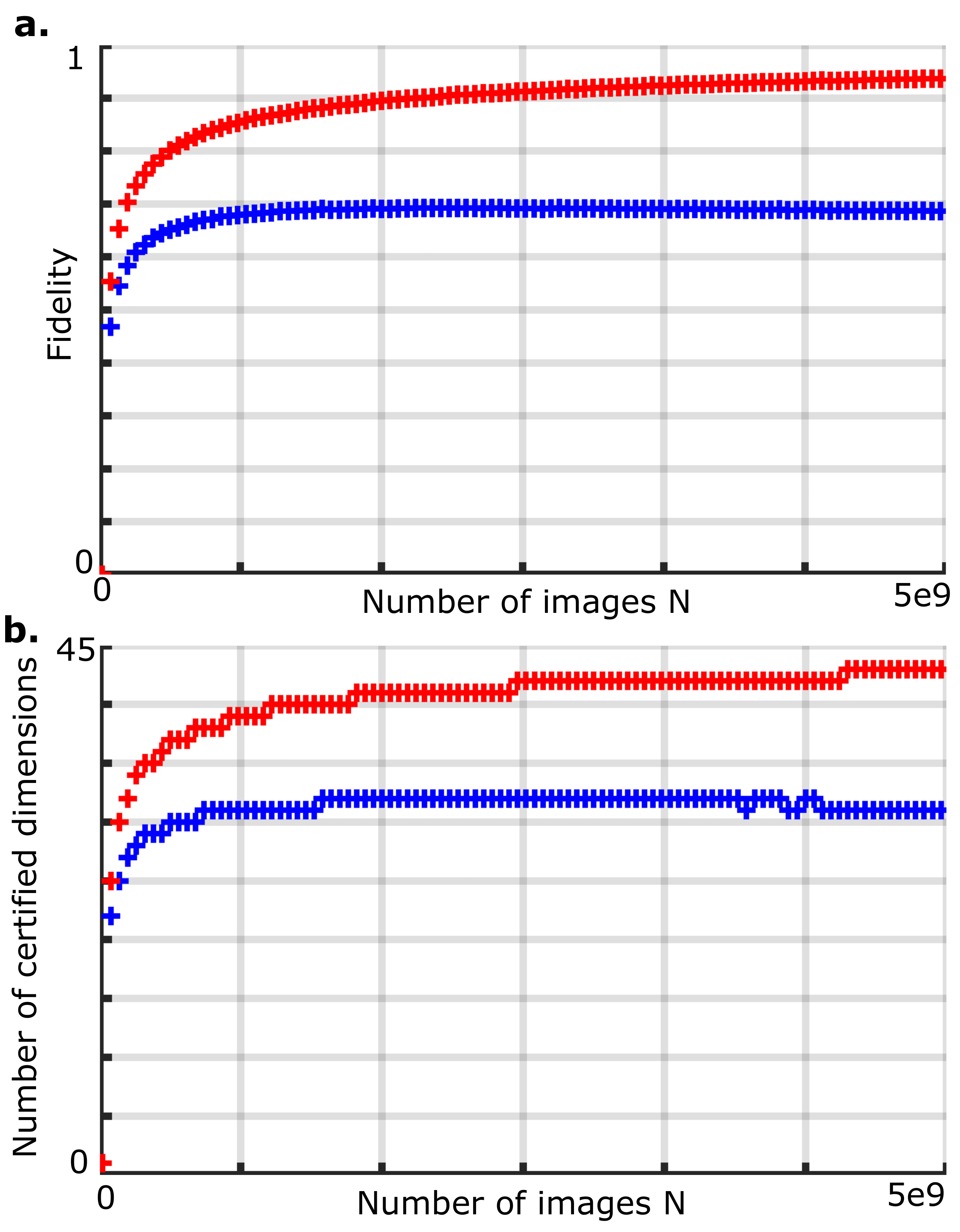} % this command will be ignored
\caption{\textbf{Simulated fidelity and number of certified dimensions in function of the number of frames.} \textbf{(a)} Fidelity in function of the number of frames $N$ in the case of a maximally-entangled state (blue) and a non-maximally-entangled state (red). \textbf{(b)} Number of certified dimensions in function of the number of frames $N$ in the case of a maximally-entangled state (blue) and a non-maximally-entangled state (red).  Simulations parameters are $\alpha = 1/45$, $\alpha' = 0.0001$ and $K = 1$.
\label{FigureSM5}}
\end{figure}

To verify the presence of a plateau when measuring the certified dimension in function of the number of frames (Fig.~\ref{Figure4}), we performed a simulation. For that, we used the model of the JPD dependency with number of frames developed by Reichert et al.~\cite{reichert_optimizing_2018}. In this model, the standard deviation of a JPD coefficient $\Gamma_{ijkl}$ computed from $N$ frames scales as $\sigma_{\Gamma}=\dfrac{1}{\sqrt{N}}$. Figures~\ref{FigureSM5}.a.b show simulation results obtained for two different quantum states, one maximally entangled (blue curves) and the second non maximally entangled (red curves). The main steps of the algorithm to obtain these results are described in the following: 
\begin{enumerate}
\item \textit{Generation of correlation matrices for a given $N$.} In the case of the maximally entangled state, a pair of correlation matrices (momentum and position) is generated using the \textit{normrnd(mu,sigma)} function of Matlab. This function generates a random number from the normal distribution with mean parameter $\mu$ and standard deviation parameter $\sigma$. Diagonal values of the two matrices are obtained using the parameters $\mu = \alpha > 0$ and $\sigma = K/\sqrt{N}$, where $K$ is a simulation parameter and $N$ is a parameter representing the number of frames. Off-diagonal values are obtained using parameters $\mu = 0$ and $\sigma = K/\sqrt{N}$. In the case of the non-maximally entangled state, the only difference if that the off-diagonal elements are obtained from a random process with a non-zero mean parameter $\mu = \alpha' < \alpha $. Matrices have dimensions $45\times45$. Figure~\ref{FigureSM4bis} shows an example of the generation of one correlation matrix.
\item \textit{Calculation of the fidelity and the number of certified dimensions.} For a given $N$, a set $100$ pairs of correlation matrices is generated in each case (i.e. maximally and non-maximally entangled). The fidelity is then calculated for each pair of matrices using equations~\eqref{F1} et~\eqref{f2tilde}, and averaged over the set. Then, equation~\eqref{Eq2} is used to calculate the number of certified dimensions. This process is repeated for different number of frames $N$ to obtain the curves in Figures~\ref{FigureSM5}.a.b.
\end{enumerate}
Results shown in Figure~\ref{FigureSM5} confirm the presence of plateaus for large $N$ in each case.

%\bibliographystyle{apsrev4-1}
%\bibliography{Biblio_2}

\begin{thebibliography}{55}%
\makeatletter
\providecommand \@ifxundefined [1]{%
 \@ifx{#1\undefined}
}%
\providecommand \@ifnum [1]{%
 \ifnum #1\expandafter \@firstoftwo
 \else \expandafter \@secondoftwo
 \fi
}%
\providecommand \@ifx [1]{%
 \ifx #1\expandafter \@firstoftwo
 \else \expandafter \@secondoftwo
 \fi
}%
\providecommand \natexlab [1]{#1}%
\providecommand \enquote  [1]{``#1''}%
\providecommand \bibnamefont  [1]{#1}%
\providecommand \bibfnamefont [1]{#1}%
\providecommand \citenamefont [1]{#1}%
\providecommand \href@noop [0]{\@secondoftwo}%
\providecommand \href [0]{\begingroup \@sanitize@url \@href}%
\providecommand \@href[1]{\@@startlink{#1}\@@href}%
\providecommand \@@href[1]{\endgroup#1\@@endlink}%
\providecommand \@sanitize@url [0]{\catcode `\\12\catcode `\$12\catcode
  `\&12\catcode `\#12\catcode `\^12\catcode `\_12\catcode `\%12\relax}%
\providecommand \@@startlink[1]{}%
\providecommand \@@endlink[0]{}%
\providecommand \url  [0]{\begingroup\@sanitize@url \@url }%
\providecommand \@url [1]{\endgroup\@href {#1}{\urlprefix }}%
\providecommand \urlprefix  [0]{URL }%
\providecommand \Eprint [0]{\href }%
\providecommand \doibase [0]{https://doi.org/}%
\providecommand \selectlanguage [0]{\@gobble}%
\providecommand \bibinfo  [0]{\@secondoftwo}%
\providecommand \bibfield  [0]{\@secondoftwo}%
\providecommand \translation [1]{[#1]}%
\providecommand \BibitemOpen [0]{}%
\providecommand \bibitemStop [0]{}%
\providecommand \bibitemNoStop [0]{.\EOS\space}%
\providecommand \EOS [0]{\spacefactor3000\relax}%
\providecommand \BibitemShut  [1]{\csname bibitem#1\endcsname}%
\let\auto@bib@innerbib\@empty
%</preamble>
\bibitem [{\citenamefont {Mirhosseini}\ \emph {et~al.}(2015)\citenamefont
  {Mirhosseini}, \citenamefont {Magaña-Loaiza}, \citenamefont {O’Sullivan},
  \citenamefont {Rodenburg}, \citenamefont {Malik}, \citenamefont {Lavery},
  \citenamefont {Padgett}, \citenamefont {Gauthier},\ and\ \citenamefont
  {Boyd}}]{mirhosseini_high-dimensional_2015}%
  \BibitemOpen
  \bibfield  {author} {\bibinfo {author} {\bibfnamefont {M.}~\bibnamefont
  {Mirhosseini}}, \bibinfo {author} {\bibfnamefont {O.~S.}\ \bibnamefont
  {Magaña-Loaiza}}, \bibinfo {author} {\bibfnamefont {M.~N.}\ \bibnamefont
  {O’Sullivan}}, \bibinfo {author} {\bibfnamefont {B.}~\bibnamefont
  {Rodenburg}}, \bibinfo {author} {\bibfnamefont {M.}~\bibnamefont {Malik}},
  \bibinfo {author} {\bibfnamefont {M.~P.~J.}\ \bibnamefont {Lavery}}, \bibinfo
  {author} {\bibfnamefont {M.~J.}\ \bibnamefont {Padgett}}, \bibinfo {author}
  {\bibfnamefont {D.~J.}\ \bibnamefont {Gauthier}},\ and\ \bibinfo {author}
  {\bibfnamefont {R.~W.}\ \bibnamefont {Boyd}},\ }\bibfield  {title}
  {{\selectlanguage {en}\bibinfo {title} {High-dimensional quantum cryptography
  with twisted light}},\ }\href {https://doi.org/10.1088/1367-2630/17/3/033033}
  {\bibfield  {journal} {\bibinfo  {journal} {New Journal of Physics}\ }\textbf
  {\bibinfo {volume} {17}},\ \bibinfo {pages} {033033} (\bibinfo {year}
  {2015})}\BibitemShut {NoStop}%
\bibitem [{\citenamefont {Ecker}\ \emph {et~al.}(2019)\citenamefont {Ecker},
  \citenamefont {Bouchard}, \citenamefont {Bulla}, \citenamefont {Brandt},
  \citenamefont {Kohout}, \citenamefont {Steinlechner}, \citenamefont
  {Fickler}, \citenamefont {Malik}, \citenamefont {Guryanova}, \citenamefont
  {Ursin},\ and\ \citenamefont {Huber}}]{ecker_overcoming_2019}%
  \BibitemOpen
  \bibfield  {author} {\bibinfo {author} {\bibfnamefont {S.}~\bibnamefont
  {Ecker}}, \bibinfo {author} {\bibfnamefont {F.}~\bibnamefont {Bouchard}},
  \bibinfo {author} {\bibfnamefont {L.}~\bibnamefont {Bulla}}, \bibinfo
  {author} {\bibfnamefont {F.}~\bibnamefont {Brandt}}, \bibinfo {author}
  {\bibfnamefont {O.}~\bibnamefont {Kohout}}, \bibinfo {author} {\bibfnamefont
  {F.}~\bibnamefont {Steinlechner}}, \bibinfo {author} {\bibfnamefont
  {R.}~\bibnamefont {Fickler}}, \bibinfo {author} {\bibfnamefont
  {M.}~\bibnamefont {Malik}}, \bibinfo {author} {\bibfnamefont
  {Y.}~\bibnamefont {Guryanova}}, \bibinfo {author} {\bibfnamefont
  {R.}~\bibnamefont {Ursin}},\ and\ \bibinfo {author} {\bibfnamefont
  {M.}~\bibnamefont {Huber}},\ }\bibfield  {title} {\bibinfo {title}
  {Overcoming {Noise} in {Entanglement} {Distribution}},\ }\href
  {https://doi.org/10.1103/PhysRevX.9.041042} {\bibfield  {journal} {\bibinfo
  {journal} {Physical Review X}\ }\textbf {\bibinfo {volume} {9}},\ \bibinfo
  {pages} {041042} (\bibinfo {year} {2019})}\BibitemShut {NoStop}%
\bibitem [{\citenamefont {Acín}\ \emph {et~al.}(2007)\citenamefont {Acín},
  \citenamefont {Brunner}, \citenamefont {Gisin}, \citenamefont {Massar},
  \citenamefont {Pironio},\ and\ \citenamefont
  {Scarani}}]{acin_device-independent_2007}%
  \BibitemOpen
  \bibfield  {author} {\bibinfo {author} {\bibfnamefont {A.}~\bibnamefont
  {Acín}}, \bibinfo {author} {\bibfnamefont {N.}~\bibnamefont {Brunner}},
  \bibinfo {author} {\bibfnamefont {N.}~\bibnamefont {Gisin}}, \bibinfo
  {author} {\bibfnamefont {S.}~\bibnamefont {Massar}}, \bibinfo {author}
  {\bibfnamefont {S.}~\bibnamefont {Pironio}},\ and\ \bibinfo {author}
  {\bibfnamefont {V.}~\bibnamefont {Scarani}},\ }\bibfield  {title} {\bibinfo
  {title} {Device-{Independent} {Security} of {Quantum} {Cryptography} against
  {Collective} {Attacks}},\ }\href
  {https://doi.org/10.1103/PhysRevLett.98.230501} {\bibfield  {journal}
  {\bibinfo  {journal} {Physical Review Letters}\ }\textbf {\bibinfo {volume}
  {98}},\ \bibinfo {pages} {230501} (\bibinfo {year} {2007})}\BibitemShut
  {NoStop}%
\bibitem [{\citenamefont {Brida}\ \emph {et~al.}(2010)\citenamefont {Brida},
  \citenamefont {Genovese},\ and\ \citenamefont
  {Berchera}}]{brida_experimental_2010}%
  \BibitemOpen
  \bibfield  {author} {\bibinfo {author} {\bibfnamefont {G.}~\bibnamefont
  {Brida}}, \bibinfo {author} {\bibfnamefont {M.}~\bibnamefont {Genovese}},\
  and\ \bibinfo {author} {\bibfnamefont {I.~R.}\ \bibnamefont {Berchera}},\
  }\bibfield  {title} {{\selectlanguage {en}\bibinfo {title} {Experimental
  realization of sub-shot-noise quantum imaging}},\ }\href
  {https://doi.org/10.1038/nphoton.2010.29} {\bibfield  {journal} {\bibinfo
  {journal} {Nature Photonics}\ }\textbf {\bibinfo {volume} {4}},\ \bibinfo
  {pages} {227} (\bibinfo {year} {2010})}\BibitemShut {NoStop}%
\bibitem [{\citenamefont {Toninelli}\ \emph {et~al.}(2019)\citenamefont
  {Toninelli}, \citenamefont {Moreau}, \citenamefont {Gregory}, \citenamefont
  {Mihalyi}, \citenamefont {Edgar}, \citenamefont {Radwell},\ and\
  \citenamefont {Padgett}}]{toninelli_resolution-enhanced_2019}%
  \BibitemOpen
  \bibfield  {author} {\bibinfo {author} {\bibfnamefont {E.}~\bibnamefont
  {Toninelli}}, \bibinfo {author} {\bibfnamefont {P.-A.}\ \bibnamefont
  {Moreau}}, \bibinfo {author} {\bibfnamefont {T.}~\bibnamefont {Gregory}},
  \bibinfo {author} {\bibfnamefont {A.}~\bibnamefont {Mihalyi}}, \bibinfo
  {author} {\bibfnamefont {M.}~\bibnamefont {Edgar}}, \bibinfo {author}
  {\bibfnamefont {N.}~\bibnamefont {Radwell}},\ and\ \bibinfo {author}
  {\bibfnamefont {M.}~\bibnamefont {Padgett}},\ }\bibfield  {title}
  {{\selectlanguage {en}\bibinfo {title} {Resolution-enhanced quantum imaging
  by centroid estimation of biphotons}},\ }\href
  {https://doi.org/10.1364/OPTICA.6.000347} {\bibfield  {journal} {\bibinfo
  {journal} {Optica}\ }\textbf {\bibinfo {volume} {6}},\ \bibinfo {pages} {347}
  (\bibinfo {year} {2019})}\BibitemShut {NoStop}%
\bibitem [{\citenamefont {Camphausen}\ \emph {et~al.}(2021)\citenamefont
  {Camphausen}, \citenamefont {Cuevas}, \citenamefont {Duempelmann},
  \citenamefont {Terborg}, \citenamefont {Wajs}, \citenamefont {Tisa},
  \citenamefont {Ruggeri}, \citenamefont {Cusini}, \citenamefont
  {Steinlechner},\ and\ \citenamefont
  {Pruneri}}]{camphausen_quantum-enhanced_2021}%
  \BibitemOpen
  \bibfield  {author} {\bibinfo {author} {\bibfnamefont {R.}~\bibnamefont
  {Camphausen}}, \bibinfo {author} {\bibfnamefont {A.}~\bibnamefont {Cuevas}},
  \bibinfo {author} {\bibfnamefont {L.}~\bibnamefont {Duempelmann}}, \bibinfo
  {author} {\bibfnamefont {R.~A.}\ \bibnamefont {Terborg}}, \bibinfo {author}
  {\bibfnamefont {E.}~\bibnamefont {Wajs}}, \bibinfo {author} {\bibfnamefont
  {S.}~\bibnamefont {Tisa}}, \bibinfo {author} {\bibfnamefont {A.}~\bibnamefont
  {Ruggeri}}, \bibinfo {author} {\bibfnamefont {I.}~\bibnamefont {Cusini}},
  \bibinfo {author} {\bibfnamefont {F.}~\bibnamefont {Steinlechner}},\ and\
  \bibinfo {author} {\bibfnamefont {V.}~\bibnamefont {Pruneri}},\ }\bibfield
  {title} {\bibinfo {title} {A quantum-enhanced wide-field phase imager},\
  }\href {https://doi.org/10.1126/sciadv.abj2155} {\bibfield  {journal}
  {\bibinfo  {journal} {Science Advances}\ }\textbf {\bibinfo {volume} {7}},\
  \bibinfo {pages} {eabj2155} (\bibinfo {year} {2021})}\BibitemShut {NoStop}%
\bibitem [{\citenamefont {Defienne}\ \emph {et~al.}(2022)\citenamefont
  {Defienne}, \citenamefont {Cameron}, \citenamefont {Ndagano}, \citenamefont
  {Lyons}, \citenamefont {Reichert}, \citenamefont {Zhao}, \citenamefont
  {Harvey}, \citenamefont {Charbon}, \citenamefont {Fleischer},\ and\
  \citenamefont {Faccio}}]{defienne_pixel_2022-2}%
  \BibitemOpen
  \bibfield  {author} {\bibinfo {author} {\bibfnamefont {H.}~\bibnamefont
  {Defienne}}, \bibinfo {author} {\bibfnamefont {P.}~\bibnamefont {Cameron}},
  \bibinfo {author} {\bibfnamefont {B.}~\bibnamefont {Ndagano}}, \bibinfo
  {author} {\bibfnamefont {A.}~\bibnamefont {Lyons}}, \bibinfo {author}
  {\bibfnamefont {M.}~\bibnamefont {Reichert}}, \bibinfo {author}
  {\bibfnamefont {J.}~\bibnamefont {Zhao}}, \bibinfo {author} {\bibfnamefont
  {A.~R.}\ \bibnamefont {Harvey}}, \bibinfo {author} {\bibfnamefont
  {E.}~\bibnamefont {Charbon}}, \bibinfo {author} {\bibfnamefont {J.~W.}\
  \bibnamefont {Fleischer}},\ and\ \bibinfo {author} {\bibfnamefont
  {D.}~\bibnamefont {Faccio}},\ }\bibfield  {title} {{\selectlanguage
  {en}\bibinfo {title} {Pixel super-resolution with spatially entangled
  photons}},\ }\href {https://doi.org/10.1038/s41467-022-31052-6} {\bibfield
  {journal} {\bibinfo  {journal} {Nature Communications}\ }\textbf {\bibinfo
  {volume} {13}},\ \bibinfo {pages} {3566} (\bibinfo {year}
  {2022})}\BibitemShut {NoStop}%
\bibitem [{\citenamefont {Ndagano}\ \emph {et~al.}(2022)\citenamefont
  {Ndagano}, \citenamefont {Defienne}, \citenamefont {Branford}, \citenamefont
  {Shah}, \citenamefont {Lyons}, \citenamefont {Westerberg}, \citenamefont
  {Gauger},\ and\ \citenamefont {Faccio}}]{ndagano_quantum_2022}%
  \BibitemOpen
  \bibfield  {author} {\bibinfo {author} {\bibfnamefont {B.}~\bibnamefont
  {Ndagano}}, \bibinfo {author} {\bibfnamefont {H.}~\bibnamefont {Defienne}},
  \bibinfo {author} {\bibfnamefont {D.}~\bibnamefont {Branford}}, \bibinfo
  {author} {\bibfnamefont {Y.~D.}\ \bibnamefont {Shah}}, \bibinfo {author}
  {\bibfnamefont {A.}~\bibnamefont {Lyons}}, \bibinfo {author} {\bibfnamefont
  {N.}~\bibnamefont {Westerberg}}, \bibinfo {author} {\bibfnamefont {E.~M.}\
  \bibnamefont {Gauger}},\ and\ \bibinfo {author} {\bibfnamefont
  {D.}~\bibnamefont {Faccio}},\ }\bibfield  {title} {\bibinfo {title} {Quantum
  microscopy based on hong--ou--mandel interference},\ }\href@noop {}
  {\bibfield  {journal} {\bibinfo  {journal} {Nature Photonics}\ }\textbf
  {\bibinfo {volume} {16}},\ \bibinfo {pages} {384} (\bibinfo {year}
  {2022})}\BibitemShut {NoStop}%
\bibitem [{\citenamefont {Defienne}\ \emph {et~al.}(2019)\citenamefont
  {Defienne}, \citenamefont {Reichert}, \citenamefont {Fleischer},\ and\
  \citenamefont {Faccio}}]{defienne_quantum_2019-1}%
  \BibitemOpen
  \bibfield  {author} {\bibinfo {author} {\bibfnamefont {H.}~\bibnamefont
  {Defienne}}, \bibinfo {author} {\bibfnamefont {M.}~\bibnamefont {Reichert}},
  \bibinfo {author} {\bibfnamefont {J.~W.}\ \bibnamefont {Fleischer}},\ and\
  \bibinfo {author} {\bibfnamefont {D.}~\bibnamefont {Faccio}},\ }\bibfield
  {title} {{\selectlanguage {en}\bibinfo {title} {Quantum image
  distillation}},\ }\href {https://doi.org/10.1126/sciadv.aax0307} {\bibfield
  {journal} {\bibinfo  {journal} {Science Advances}\ }\textbf {\bibinfo
  {volume} {5}},\ \bibinfo {pages} {eaax0307} (\bibinfo {year}
  {2019})}\BibitemShut {NoStop}%
\bibitem [{\citenamefont {Gregory}\ \emph {et~al.}(2020)\citenamefont
  {Gregory}, \citenamefont {Moreau}, \citenamefont {Toninelli},\ and\
  \citenamefont {Padgett}}]{gregory_imaging_2020}%
  \BibitemOpen
  \bibfield  {author} {\bibinfo {author} {\bibfnamefont {T.}~\bibnamefont
  {Gregory}}, \bibinfo {author} {\bibfnamefont {P.-A.}\ \bibnamefont {Moreau}},
  \bibinfo {author} {\bibfnamefont {E.}~\bibnamefont {Toninelli}},\ and\
  \bibinfo {author} {\bibfnamefont {M.~J.}\ \bibnamefont {Padgett}},\
  }\bibfield  {title} {{\selectlanguage {en}\bibinfo {title} {Imaging through
  noise with quantum illumination}},\ }\href
  {https://doi.org/10.1126/sciadv.aay2652} {\bibfield  {journal} {\bibinfo
  {journal} {Science Advances}\ }\textbf {\bibinfo {volume} {6}},\ \bibinfo
  {pages} {eaay2652} (\bibinfo {year} {2020})}\BibitemShut {NoStop}%
\bibitem [{\citenamefont {Devaux}\ \emph {et~al.}(2019)\citenamefont {Devaux},
  \citenamefont {Mosset}, \citenamefont {Bassignot},\ and\ \citenamefont
  {Lantz}}]{devaux_quantum_2019}%
  \BibitemOpen
  \bibfield  {author} {\bibinfo {author} {\bibfnamefont {F.}~\bibnamefont
  {Devaux}}, \bibinfo {author} {\bibfnamefont {A.}~\bibnamefont {Mosset}},
  \bibinfo {author} {\bibfnamefont {F.}~\bibnamefont {Bassignot}},\ and\
  \bibinfo {author} {\bibfnamefont {E.}~\bibnamefont {Lantz}},\ }\bibfield
  {title} {\bibinfo {title} {Quantum holography with biphotons of high
  {Schmidt} number},\ }\href {https://doi.org/10.1103/PhysRevA.99.033854}
  {\bibfield  {journal} {\bibinfo  {journal} {Physical Review A}\ }\textbf
  {\bibinfo {volume} {99}},\ \bibinfo {pages} {033854} (\bibinfo {year}
  {2019})}\BibitemShut {NoStop}%
\bibitem [{\citenamefont {Defienne}\ \emph {et~al.}(2021)\citenamefont
  {Defienne}, \citenamefont {Ndagano}, \citenamefont {Lyons},\ and\
  \citenamefont {Faccio}}]{defienne_polarization_2021-4}%
  \BibitemOpen
  \bibfield  {author} {\bibinfo {author} {\bibfnamefont {H.}~\bibnamefont
  {Defienne}}, \bibinfo {author} {\bibfnamefont {B.}~\bibnamefont {Ndagano}},
  \bibinfo {author} {\bibfnamefont {A.}~\bibnamefont {Lyons}},\ and\ \bibinfo
  {author} {\bibfnamefont {D.}~\bibnamefont {Faccio}},\ }\bibfield  {title}
  {{\selectlanguage {en}\bibinfo {title} {Polarization entanglement-enabled
  quantum holography}},\ }\href {https://doi.org/10.1038/s41567-020-01156-1}
  {\bibfield  {journal} {\bibinfo  {journal} {Nature Physics}\ }\textbf
  {\bibinfo {volume} {17}},\ \bibinfo {pages} {591} (\bibinfo {year}
  {2021})}\BibitemShut {NoStop}%
\bibitem [{\citenamefont {Vellekoop}\ and\ \citenamefont
  {Mosk}(2007)}]{vellekoop_focusing_2007}%
  \BibitemOpen
  \bibfield  {author} {\bibinfo {author} {\bibfnamefont {I.~M.}\ \bibnamefont
  {Vellekoop}}\ and\ \bibinfo {author} {\bibfnamefont {A.~P.}\ \bibnamefont
  {Mosk}},\ }\bibfield  {title} {\bibinfo {title} {Focusing coherent light
  through opaque strongly scattering media},\ }\href
  {http://www.opticsinfobase.org/abstract.cfm?uri=ol-32-16-2309} {\bibfield
  {journal} {\bibinfo  {journal} {Optics letters}\ }\textbf {\bibinfo {volume}
  {32}},\ \bibinfo {pages} {2309} (\bibinfo {year} {2007})}\BibitemShut
  {NoStop}%
\bibitem [{\citenamefont {Popoff}\ \emph {et~al.}(2010)\citenamefont {Popoff},
  \citenamefont {Lerosey}, \citenamefont {Carminati}, \citenamefont {Fink},
  \citenamefont {Boccara},\ and\ \citenamefont
  {Gigan}}]{popoff_measuring_2010}%
  \BibitemOpen
  \bibfield  {author} {\bibinfo {author} {\bibfnamefont {S.~M.}\ \bibnamefont
  {Popoff}}, \bibinfo {author} {\bibfnamefont {G.}~\bibnamefont {Lerosey}},
  \bibinfo {author} {\bibfnamefont {R.}~\bibnamefont {Carminati}}, \bibinfo
  {author} {\bibfnamefont {M.}~\bibnamefont {Fink}}, \bibinfo {author}
  {\bibfnamefont {A.~C.}\ \bibnamefont {Boccara}},\ and\ \bibinfo {author}
  {\bibfnamefont {S.}~\bibnamefont {Gigan}},\ }\bibfield  {title} {\bibinfo
  {title} {Measuring the transmission matrix in optics: an approach to the
  study and control of light propagation in disordered media},\ }\href
  {http://journals.aps.org/prl/abstract/10.1103/PhysRevLett.104.100601}
  {\bibfield  {journal} {\bibinfo  {journal} {Physical review letters}\
  }\textbf {\bibinfo {volume} {104}},\ \bibinfo {pages} {100601} (\bibinfo
  {year} {2010})}\BibitemShut {NoStop}%
\bibitem [{\citenamefont {Carpenter}\ \emph {et~al.}(2014)\citenamefont
  {Carpenter}, \citenamefont {Eggleton},\ and\ \citenamefont
  {Schröder}}]{carpenter_110x110_2014}%
  \BibitemOpen
  \bibfield  {author} {\bibinfo {author} {\bibfnamefont {J.}~\bibnamefont
  {Carpenter}}, \bibinfo {author} {\bibfnamefont {B.~J.}\ \bibnamefont
  {Eggleton}},\ and\ \bibinfo {author} {\bibfnamefont {J.}~\bibnamefont
  {Schröder}},\ }\bibfield  {title} {\bibinfo {title} {110x110 optical mode
  transfer matrix inversion},\ }\href {https://doi.org/10.1364/OE.22.000096}
  {\bibfield  {journal} {\bibinfo  {journal} {Optics Express}\ }\textbf
  {\bibinfo {volume} {22}},\ \bibinfo {pages} {96} (\bibinfo {year}
  {2014})}\BibitemShut {NoStop}%
\bibitem [{\citenamefont {Plöschner}\ \emph {et~al.}(2015)\citenamefont
  {Plöschner}, \citenamefont {Tyc},\ and\ \citenamefont
  {Čižmár}}]{ploschner_seeing_2015-3}%
  \BibitemOpen
  \bibfield  {author} {\bibinfo {author} {\bibfnamefont {M.}~\bibnamefont
  {Plöschner}}, \bibinfo {author} {\bibfnamefont {T.}~\bibnamefont {Tyc}},\
  and\ \bibinfo {author} {\bibfnamefont {T.}~\bibnamefont {Čižmár}},\
  }\bibfield  {title} {{\selectlanguage {en}\bibinfo {title} {Seeing through
  chaos in multimode fibres}},\ }\href
  {https://doi.org/10.1038/nphoton.2015.112} {\bibfield  {journal} {\bibinfo
  {journal} {Nature Photonics}\ }\textbf {\bibinfo {volume} {9}},\ \bibinfo
  {pages} {529} (\bibinfo {year} {2015})}\BibitemShut {NoStop}%
\bibitem [{\citenamefont {Kang}\ \emph {et~al.}(2015)\citenamefont {Kang},
  \citenamefont {Jeong}, \citenamefont {Choi}, \citenamefont {Ko},
  \citenamefont {Yang}, \citenamefont {Joo}, \citenamefont {Lee}, \citenamefont
  {Lim}, \citenamefont {Park},\ and\ \citenamefont {Choi}}]{kang_imaging_2015}%
  \BibitemOpen
  \bibfield  {author} {\bibinfo {author} {\bibfnamefont {S.}~\bibnamefont
  {Kang}}, \bibinfo {author} {\bibfnamefont {S.}~\bibnamefont {Jeong}},
  \bibinfo {author} {\bibfnamefont {W.}~\bibnamefont {Choi}}, \bibinfo {author}
  {\bibfnamefont {H.}~\bibnamefont {Ko}}, \bibinfo {author} {\bibfnamefont
  {T.~D.}\ \bibnamefont {Yang}}, \bibinfo {author} {\bibfnamefont {J.~H.}\
  \bibnamefont {Joo}}, \bibinfo {author} {\bibfnamefont {J.-S.}\ \bibnamefont
  {Lee}}, \bibinfo {author} {\bibfnamefont {Y.-S.}\ \bibnamefont {Lim}},
  \bibinfo {author} {\bibfnamefont {Q.-H.}\ \bibnamefont {Park}},\ and\
  \bibinfo {author} {\bibfnamefont {W.}~\bibnamefont {Choi}},\ }\bibfield
  {title} {{\selectlanguage {en}\bibinfo {title} {Imaging deep within a
  scattering medium using collective accumulation of single-scattered waves}},\
  }\href {https://doi.org/10.1038/nphoton.2015.24} {\bibfield  {journal}
  {\bibinfo  {journal} {Nature Photonics}\ }\textbf {\bibinfo {volume} {9}},\
  \bibinfo {pages} {253} (\bibinfo {year} {2015})}\BibitemShut {NoStop}%
\bibitem [{\citenamefont {Badon}\ \emph {et~al.}(2016)\citenamefont {Badon},
  \citenamefont {Li}, \citenamefont {Lerosey}, \citenamefont {Boccara},
  \citenamefont {Fink},\ and\ \citenamefont {Aubry}}]{badon_smart_2016}%
  \BibitemOpen
  \bibfield  {author} {\bibinfo {author} {\bibfnamefont {A.}~\bibnamefont
  {Badon}}, \bibinfo {author} {\bibfnamefont {D.}~\bibnamefont {Li}}, \bibinfo
  {author} {\bibfnamefont {G.}~\bibnamefont {Lerosey}}, \bibinfo {author}
  {\bibfnamefont {A.~C.}\ \bibnamefont {Boccara}}, \bibinfo {author}
  {\bibfnamefont {M.}~\bibnamefont {Fink}},\ and\ \bibinfo {author}
  {\bibfnamefont {A.}~\bibnamefont {Aubry}},\ }\bibfield  {title}
  {{\selectlanguage {en}\bibinfo {title} {Smart optical coherence tomography
  for ultra-deep imaging through highly scattering media}},\ }\href
  {https://doi.org/10.1126/sciadv.1600370} {\bibfield  {journal} {\bibinfo
  {journal} {Science Advances}\ }\textbf {\bibinfo {volume} {2}},\ \bibinfo
  {pages} {e1600370} (\bibinfo {year} {2016})}\BibitemShut {NoStop}%
\bibitem [{\citenamefont {Carpenter}\ \emph {et~al.}(2013)\citenamefont
  {Carpenter}, \citenamefont {Xiong}, \citenamefont {Collins}, \citenamefont
  {Li}, \citenamefont {Krauss}, \citenamefont {Eggleton}, \citenamefont
  {Clark},\ and\ \citenamefont {Schröder}}]{carpenter_mode_2013}%
  \BibitemOpen
  \bibfield  {author} {\bibinfo {author} {\bibfnamefont {J.}~\bibnamefont
  {Carpenter}}, \bibinfo {author} {\bibfnamefont {C.}~\bibnamefont {Xiong}},
  \bibinfo {author} {\bibfnamefont {M.~J.}\ \bibnamefont {Collins}}, \bibinfo
  {author} {\bibfnamefont {J.}~\bibnamefont {Li}}, \bibinfo {author}
  {\bibfnamefont {T.~F.}\ \bibnamefont {Krauss}}, \bibinfo {author}
  {\bibfnamefont {B.~J.}\ \bibnamefont {Eggleton}}, \bibinfo {author}
  {\bibfnamefont {A.~S.}\ \bibnamefont {Clark}},\ and\ \bibinfo {author}
  {\bibfnamefont {J.}~\bibnamefont {Schröder}},\ }\bibfield  {title} {\bibinfo
  {title} {Mode multiplexed single-photon and classical channels in a few-mode
  fiber},\ }\href {https://doi.org/10.1364/OE.21.028794} {\bibfield  {journal}
  {\bibinfo  {journal} {Optics Express}\ }\textbf {\bibinfo {volume} {21}},\
  \bibinfo {pages} {28794} (\bibinfo {year} {2013})}\BibitemShut {NoStop}%
\bibitem [{\citenamefont {Huisman}\ \emph {et~al.}(2014)\citenamefont
  {Huisman}, \citenamefont {Huisman}, \citenamefont {Mosk},\ and\ \citenamefont
  {Pinkse}}]{huisman_controlling_2014}%
  \BibitemOpen
  \bibfield  {author} {\bibinfo {author} {\bibfnamefont {T.~J.}\ \bibnamefont
  {Huisman}}, \bibinfo {author} {\bibfnamefont {S.~R.}\ \bibnamefont
  {Huisman}}, \bibinfo {author} {\bibfnamefont {A.~P.}\ \bibnamefont {Mosk}},\
  and\ \bibinfo {author} {\bibfnamefont {P.~W.~H.}\ \bibnamefont {Pinkse}},\
  }\bibfield  {title} {{\selectlanguage {en}\bibinfo {title} {Controlling
  single-photon {Fock}-state propagation through opaque scattering media}},\
  }\href {https://doi.org/10.1007/s00340-013-5742-5} {\bibfield  {journal}
  {\bibinfo  {journal} {Applied Physics B}\ }\textbf {\bibinfo {volume}
  {116}},\ \bibinfo {pages} {603} (\bibinfo {year} {2014})}\BibitemShut
  {NoStop}%
\bibitem [{\citenamefont {Defienne}\ \emph {et~al.}(2014)\citenamefont
  {Defienne}, \citenamefont {Barbieri}, \citenamefont {Chalopin}, \citenamefont
  {Chatel}, \citenamefont {Walmsley}, \citenamefont {Smith},\ and\
  \citenamefont {Gigan}}]{defienne_nonclassical_2014}%
  \BibitemOpen
  \bibfield  {author} {\bibinfo {author} {\bibfnamefont {H.}~\bibnamefont
  {Defienne}}, \bibinfo {author} {\bibfnamefont {M.}~\bibnamefont {Barbieri}},
  \bibinfo {author} {\bibfnamefont {B.}~\bibnamefont {Chalopin}}, \bibinfo
  {author} {\bibfnamefont {B.}~\bibnamefont {Chatel}}, \bibinfo {author}
  {\bibfnamefont {I.~A.}\ \bibnamefont {Walmsley}}, \bibinfo {author}
  {\bibfnamefont {B.~J.}\ \bibnamefont {Smith}},\ and\ \bibinfo {author}
  {\bibfnamefont {S.}~\bibnamefont {Gigan}},\ }\bibfield  {title} {\bibinfo
  {title} {Nonclassical light manipulation in a multiple-scattering medium},\
  }\href {https://doi.org/10.1364/OL.39.006090} {\bibfield  {journal} {\bibinfo
   {journal} {Optics Letters}\ }\textbf {\bibinfo {volume} {39}},\ \bibinfo
  {pages} {6090} (\bibinfo {year} {2014})}\BibitemShut {NoStop}%
\bibitem [{\citenamefont {Defienne}\ \emph {et~al.}(2016)\citenamefont
  {Defienne}, \citenamefont {Barbieri}, \citenamefont {Walmsley}, \citenamefont
  {Smith},\ and\ \citenamefont {Gigan}}]{defienne_two-photon_2016}%
  \BibitemOpen
  \bibfield  {author} {\bibinfo {author} {\bibfnamefont {H.}~\bibnamefont
  {Defienne}}, \bibinfo {author} {\bibfnamefont {M.}~\bibnamefont {Barbieri}},
  \bibinfo {author} {\bibfnamefont {I.~A.}\ \bibnamefont {Walmsley}}, \bibinfo
  {author} {\bibfnamefont {B.~J.}\ \bibnamefont {Smith}},\ and\ \bibinfo
  {author} {\bibfnamefont {S.}~\bibnamefont {Gigan}},\ }\bibfield  {title}
  {{\selectlanguage {en}\bibinfo {title} {Two-photon quantum walk in a
  multimode fiber}},\ }\href {https://doi.org/10.1126/sciadv.1501054}
  {\bibfield  {journal} {\bibinfo  {journal} {Science Advances}\ }\textbf
  {\bibinfo {volume} {2}},\ \bibinfo {pages} {e1501054} (\bibinfo {year}
  {2016})}\BibitemShut {NoStop}%
\bibitem [{\citenamefont {Wolterink}\ \emph {et~al.}(2016)\citenamefont
  {Wolterink}, \citenamefont {Uppu}, \citenamefont {Ctistis}, \citenamefont
  {Vos}, \citenamefont {Boller},\ and\ \citenamefont
  {Pinkse}}]{wolterink_programmable_2016}%
  \BibitemOpen
  \bibfield  {author} {\bibinfo {author} {\bibfnamefont {T.~A.~W.}\
  \bibnamefont {Wolterink}}, \bibinfo {author} {\bibfnamefont {R.}~\bibnamefont
  {Uppu}}, \bibinfo {author} {\bibfnamefont {G.}~\bibnamefont {Ctistis}},
  \bibinfo {author} {\bibfnamefont {W.~L.}\ \bibnamefont {Vos}}, \bibinfo
  {author} {\bibfnamefont {K.-J.}\ \bibnamefont {Boller}},\ and\ \bibinfo
  {author} {\bibfnamefont {P.~W.~H.}\ \bibnamefont {Pinkse}},\ }\bibfield
  {title} {\bibinfo {title} {Programmable two-photon quantum interference in
  \$\{10\}{\textasciicircum}\{3\}\$ channels in opaque scattering media},\
  }\href {https://doi.org/10.1103/PhysRevA.93.053817} {\bibfield  {journal}
  {\bibinfo  {journal} {Physical Review A}\ }\textbf {\bibinfo {volume} {93}},\
  \bibinfo {pages} {053817} (\bibinfo {year} {2016})}\BibitemShut {NoStop}%
\bibitem [{\citenamefont {Defienne}\ \emph
  {et~al.}(2018{\natexlab{a}})\citenamefont {Defienne}, \citenamefont
  {Reichert},\ and\ \citenamefont {Fleischer}}]{defienne_adaptive_2018-3}%
  \BibitemOpen
  \bibfield  {author} {\bibinfo {author} {\bibfnamefont {H.}~\bibnamefont
  {Defienne}}, \bibinfo {author} {\bibfnamefont {M.}~\bibnamefont {Reichert}},\
  and\ \bibinfo {author} {\bibfnamefont {J.~W.}\ \bibnamefont {Fleischer}},\
  }\bibfield  {title} {\bibinfo {title} {Adaptive {Quantum} {Optics} with
  {Spatially} {Entangled} {Photon} {Pairs}},\ }\href
  {https://doi.org/10.1103/PhysRevLett.121.233601} {\bibfield  {journal}
  {\bibinfo  {journal} {Physical Review Letters}\ }\textbf {\bibinfo {volume}
  {121}},\ \bibinfo {pages} {233601} (\bibinfo {year}
  {2018}{\natexlab{a}})}\BibitemShut {NoStop}%
\bibitem [{\citenamefont {Devaux}\ \emph {et~al.}(2022)\citenamefont {Devaux},
  \citenamefont {Mosset}, \citenamefont {Popoff},\ and\ \citenamefont
  {Lantz}}]{devaux_restoring_2022}%
  \BibitemOpen
  \bibfield  {author} {\bibinfo {author} {\bibfnamefont {F.}~\bibnamefont
  {Devaux}}, \bibinfo {author} {\bibfnamefont {A.}~\bibnamefont {Mosset}},
  \bibinfo {author} {\bibfnamefont {S.~M.}\ \bibnamefont {Popoff}},\ and\
  \bibinfo {author} {\bibfnamefont {E.}~\bibnamefont {Lantz}},\ }\bibfield
  {title} {\bibinfo {title} {Restoring and tailoring very high dimensional
  spatial entanglement of a biphoton state transmitted through a scattering
  medium},\ }\bibfield  {journal} {\bibinfo  {journal} {arXiv:2206.00299
  [physics]}\ }\href {https://doi.org/10.48550/arXiv.2206.00299}
  {10.48550/arXiv.2206.00299} (\bibinfo {year} {2022})\BibitemShut {NoStop}%
\bibitem [{\citenamefont {Lib}\ \emph {et~al.}(2020)\citenamefont {Lib},
  \citenamefont {Hasson},\ and\ \citenamefont
  {Bromberg}}]{lib_real-time_2020-1}%
  \BibitemOpen
  \bibfield  {author} {\bibinfo {author} {\bibfnamefont {O.}~\bibnamefont
  {Lib}}, \bibinfo {author} {\bibfnamefont {G.}~\bibnamefont {Hasson}},\ and\
  \bibinfo {author} {\bibfnamefont {Y.}~\bibnamefont {Bromberg}},\ }\bibfield
  {title} {{\selectlanguage {en}\bibinfo {title} {Real-time shaping of
  entangled photons by classical control and feedback}},\ }\href
  {https://doi.org/10.1126/sciadv.abb6298} {\bibfield  {journal} {\bibinfo
  {journal} {Science Advances}\ }\textbf {\bibinfo {volume} {6}},\ \bibinfo
  {pages} {eabb6298} (\bibinfo {year} {2020})}\BibitemShut {NoStop}%
\bibitem [{\citenamefont {Valencia}\ \emph {et~al.}(2020)\citenamefont
  {Valencia}, \citenamefont {Goel}, \citenamefont {McCutcheon}, \citenamefont
  {Defienne},\ and\ \citenamefont {Malik}}]{valencia_unscrambling_2020-1}%
  \BibitemOpen
  \bibfield  {author} {\bibinfo {author} {\bibfnamefont {N.~H.}\ \bibnamefont
  {Valencia}}, \bibinfo {author} {\bibfnamefont {S.}~\bibnamefont {Goel}},
  \bibinfo {author} {\bibfnamefont {W.}~\bibnamefont {McCutcheon}}, \bibinfo
  {author} {\bibfnamefont {H.}~\bibnamefont {Defienne}},\ and\ \bibinfo
  {author} {\bibfnamefont {M.}~\bibnamefont {Malik}},\ }\bibfield  {title}
  {{\selectlanguage {en}\bibinfo {title} {Unscrambling entanglement through a
  complex medium}},\ }\href {https://doi.org/10.1038/s41567-020-0970-1}
  {\bibfield  {journal} {\bibinfo  {journal} {Nature Physics}\ }\textbf
  {\bibinfo {volume} {16}},\ \bibinfo {pages} {1112} (\bibinfo {year}
  {2020})}\BibitemShut {NoStop}%
\bibitem [{\citenamefont {Friis}\ \emph {et~al.}(2019)\citenamefont {Friis},
  \citenamefont {Vitagliano}, \citenamefont {Malik},\ and\ \citenamefont
  {Huber}}]{friis_entanglement_2019}%
  \BibitemOpen
  \bibfield  {author} {\bibinfo {author} {\bibfnamefont {N.}~\bibnamefont
  {Friis}}, \bibinfo {author} {\bibfnamefont {G.}~\bibnamefont {Vitagliano}},
  \bibinfo {author} {\bibfnamefont {M.}~\bibnamefont {Malik}},\ and\ \bibinfo
  {author} {\bibfnamefont {M.}~\bibnamefont {Huber}},\ }\bibfield  {title}
  {{\selectlanguage {en}\bibinfo {title} {Entanglement certification from
  theory to experiment}},\ }\href {https://doi.org/10.1038/s42254-018-0003-5}
  {\bibfield  {journal} {\bibinfo  {journal} {Nature Reviews Physics}\ }\textbf
  {\bibinfo {volume} {1}},\ \bibinfo {pages} {72} (\bibinfo {year}
  {2019})}\BibitemShut {NoStop}%
\bibitem [{\citenamefont {Ndagano}\ \emph {et~al.}(2020)\citenamefont
  {Ndagano}, \citenamefont {Defienne}, \citenamefont {Lyons}, \citenamefont
  {Starshynov}, \citenamefont {Villa}, \citenamefont {Tisa},\ and\
  \citenamefont {Faccio}}]{ndagano_imaging_2020-1}%
  \BibitemOpen
  \bibfield  {author} {\bibinfo {author} {\bibfnamefont {B.}~\bibnamefont
  {Ndagano}}, \bibinfo {author} {\bibfnamefont {H.}~\bibnamefont {Defienne}},
  \bibinfo {author} {\bibfnamefont {A.}~\bibnamefont {Lyons}}, \bibinfo
  {author} {\bibfnamefont {I.}~\bibnamefont {Starshynov}}, \bibinfo {author}
  {\bibfnamefont {F.}~\bibnamefont {Villa}}, \bibinfo {author} {\bibfnamefont
  {S.}~\bibnamefont {Tisa}},\ and\ \bibinfo {author} {\bibfnamefont
  {D.}~\bibnamefont {Faccio}},\ }\bibfield  {title} {{\selectlanguage
  {en}\bibinfo {title} {Imaging and certifying high-dimensional entanglement
  with a single-photon avalanche diode camera}},\ }\href
  {https://doi.org/10.1038/s41534-020-00324-8} {\bibfield  {journal} {\bibinfo
  {journal} {npj Quantum Information}\ }\textbf {\bibinfo {volume} {6}},\
  \bibinfo {pages} {1} (\bibinfo {year} {2020})}\BibitemShut {NoStop}%
\bibitem [{\citenamefont {Einstein}\ \emph {et~al.}(1935)\citenamefont
  {Einstein}, \citenamefont {Podolsky},\ and\ \citenamefont
  {Rosen}}]{einstein_can_1935}%
  \BibitemOpen
  \bibfield  {author} {\bibinfo {author} {\bibfnamefont {A.}~\bibnamefont
  {Einstein}}, \bibinfo {author} {\bibfnamefont {B.}~\bibnamefont {Podolsky}},\
  and\ \bibinfo {author} {\bibfnamefont {N.}~\bibnamefont {Rosen}},\ }\bibfield
   {title} {\bibinfo {title} {Can {Quantum}-{Mechanical} {Description} of
  {Physical} {Reality} {Be} {Considered} {Complete}?},\ }\href
  {https://doi.org/10.1103/PhysRev.47.777} {\bibfield  {journal} {\bibinfo
  {journal} {Physical Review}\ }\textbf {\bibinfo {volume} {47}},\ \bibinfo
  {pages} {777} (\bibinfo {year} {1935})}\BibitemShut {NoStop}%
\bibitem [{\citenamefont {Giovannetti}\ \emph {et~al.}(2003)\citenamefont
  {Giovannetti}, \citenamefont {Mancini}, \citenamefont {Vitali},\ and\
  \citenamefont {Tombesi}}]{giovannetti_characterizing_2003}%
  \BibitemOpen
  \bibfield  {author} {\bibinfo {author} {\bibfnamefont {V.}~\bibnamefont
  {Giovannetti}}, \bibinfo {author} {\bibfnamefont {S.}~\bibnamefont
  {Mancini}}, \bibinfo {author} {\bibfnamefont {D.}~\bibnamefont {Vitali}},\
  and\ \bibinfo {author} {\bibfnamefont {P.}~\bibnamefont {Tombesi}},\
  }\bibfield  {title} {\bibinfo {title} {Characterizing the entanglement of
  bipartite quantum systems},\ }\href
  {https://doi.org/10.1103/PhysRevA.67.022320} {\bibfield  {journal} {\bibinfo
  {journal} {Physical Review A}\ }\textbf {\bibinfo {volume} {67}},\ \bibinfo
  {pages} {022320} (\bibinfo {year} {2003})}\BibitemShut {NoStop}%
\bibitem [{\citenamefont {Bavaresco}\ \emph {et~al.}(2018)\citenamefont
  {Bavaresco}, \citenamefont {Valencia}, \citenamefont {Klöckl}, \citenamefont
  {Pivoluska}, \citenamefont {Erker}, \citenamefont {Friis}, \citenamefont
  {Malik},\ and\ \citenamefont {Huber}}]{bavaresco_measurements_2018}%
  \BibitemOpen
  \bibfield  {author} {\bibinfo {author} {\bibfnamefont {J.}~\bibnamefont
  {Bavaresco}}, \bibinfo {author} {\bibfnamefont {N.~H.}\ \bibnamefont
  {Valencia}}, \bibinfo {author} {\bibfnamefont {C.}~\bibnamefont {Klöckl}},
  \bibinfo {author} {\bibfnamefont {M.}~\bibnamefont {Pivoluska}}, \bibinfo
  {author} {\bibfnamefont {P.}~\bibnamefont {Erker}}, \bibinfo {author}
  {\bibfnamefont {N.}~\bibnamefont {Friis}}, \bibinfo {author} {\bibfnamefont
  {M.}~\bibnamefont {Malik}},\ and\ \bibinfo {author} {\bibfnamefont
  {M.}~\bibnamefont {Huber}},\ }\bibfield  {title} {{\selectlanguage
  {en}\bibinfo {title} {Measurements in two bases are sufficient for certifying
  high-dimensional entanglement}},\ }\href
  {https://doi.org/10.1038/s41567-018-0203-z} {\bibfield  {journal} {\bibinfo
  {journal} {Nature Physics}\ }\textbf {\bibinfo {volume} {14}},\ \bibinfo
  {pages} {1032} (\bibinfo {year} {2018})}\BibitemShut {NoStop}%
\bibitem [{\citenamefont {Bronzi}\ \emph {et~al.}(2014)\citenamefont {Bronzi},
  \citenamefont {Villa}, \citenamefont {Tisa}, \citenamefont {Tosi},
  \citenamefont {Zappa}, \citenamefont {Durini}, \citenamefont {Weyers},\ and\
  \citenamefont {Brockherde}}]{bronzi_100_2014}%
  \BibitemOpen
  \bibfield  {author} {\bibinfo {author} {\bibfnamefont {D.}~\bibnamefont
  {Bronzi}}, \bibinfo {author} {\bibfnamefont {F.}~\bibnamefont {Villa}},
  \bibinfo {author} {\bibfnamefont {S.}~\bibnamefont {Tisa}}, \bibinfo {author}
  {\bibfnamefont {A.}~\bibnamefont {Tosi}}, \bibinfo {author} {\bibfnamefont
  {F.}~\bibnamefont {Zappa}}, \bibinfo {author} {\bibfnamefont
  {D.}~\bibnamefont {Durini}}, \bibinfo {author} {\bibfnamefont
  {S.}~\bibnamefont {Weyers}},\ and\ \bibinfo {author} {\bibfnamefont
  {W.}~\bibnamefont {Brockherde}},\ }\bibfield  {title} {\bibinfo {title} {100
  000 {Frames}/s 64 x 32 {Single}-{Photon} {Detector} {Array} for 2-{D}
  {Imaging} and 3-{D} {Ranging}},\ }\href
  {https://doi.org/10.1109/JSTQE.2014.2341562} {\bibfield  {journal} {\bibinfo
  {journal} {IEEE Journal of Selected Topics in Quantum Electronics}\ }\textbf
  {\bibinfo {volume} {20}},\ \bibinfo {pages} {354} (\bibinfo {year}
  {2014})}\BibitemShut {NoStop}%
\bibitem [{\citenamefont {Erker}\ \emph {et~al.}(2017)\citenamefont {Erker},
  \citenamefont {Krenn},\ and\ \citenamefont {Huber}}]{erker_quantifying_2017}%
  \BibitemOpen
  \bibfield  {author} {\bibinfo {author} {\bibfnamefont {P.}~\bibnamefont
  {Erker}}, \bibinfo {author} {\bibfnamefont {M.}~\bibnamefont {Krenn}},\ and\
  \bibinfo {author} {\bibfnamefont {M.}~\bibnamefont {Huber}},\ }\bibfield
  {title} {{\selectlanguage {en}\bibinfo {title} {Quantifying high dimensional
  entanglement with two mutually unbiased bases}},\ }\href
  {https://doi.org/10.22331/q-2017-07-28-22} {\bibfield  {journal} {\bibinfo
  {journal} {Quantum}\ }\textbf {\bibinfo {volume} {1}},\ \bibinfo {pages} {22}
  (\bibinfo {year} {2017})}\BibitemShut {NoStop}%
\bibitem [{\citenamefont {Howell}\ \emph {et~al.}(2004)\citenamefont {Howell},
  \citenamefont {Bennink}, \citenamefont {Bentley},\ and\ \citenamefont
  {Boyd}}]{howell_realization_2004}%
  \BibitemOpen
  \bibfield  {author} {\bibinfo {author} {\bibfnamefont {J.~C.}\ \bibnamefont
  {Howell}}, \bibinfo {author} {\bibfnamefont {R.~S.}\ \bibnamefont {Bennink}},
  \bibinfo {author} {\bibfnamefont {S.~J.}\ \bibnamefont {Bentley}},\ and\
  \bibinfo {author} {\bibfnamefont {R.~W.}\ \bibnamefont {Boyd}},\ }\bibfield
  {title} {\bibinfo {title} {Realization of the {Einstein}-{Podolsky}-{Rosen}
  {Paradox} {Using} {Momentum}- and {Position}-{Entangled} {Photons} from
  {Spontaneous} {Parametric} {Down} {Conversion}},\ }\href
  {https://doi.org/10.1103/PhysRevLett.92.210403} {\bibfield  {journal}
  {\bibinfo  {journal} {Physical Review Letters}\ }\textbf {\bibinfo {volume}
  {92}},\ \bibinfo {pages} {210403} (\bibinfo {year} {2004})}\BibitemShut
  {NoStop}%
\bibitem [{\citenamefont {Moreau}\ \emph {et~al.}(2012)\citenamefont {Moreau},
  \citenamefont {Mougin-Sisini}, \citenamefont {Devaux},\ and\ \citenamefont
  {Lantz}}]{moreau_realization_2012}%
  \BibitemOpen
  \bibfield  {author} {\bibinfo {author} {\bibfnamefont {P.-A.}\ \bibnamefont
  {Moreau}}, \bibinfo {author} {\bibfnamefont {J.}~\bibnamefont
  {Mougin-Sisini}}, \bibinfo {author} {\bibfnamefont {F.}~\bibnamefont
  {Devaux}},\ and\ \bibinfo {author} {\bibfnamefont {E.}~\bibnamefont
  {Lantz}},\ }\bibfield  {title} {\bibinfo {title} {Realization of the purely
  spatial {Einstein}-{Podolsky}-{Rosen} paradox in full-field images of
  spontaneous parametric down-conversion},\ }\href
  {https://doi.org/10.1103/PhysRevA.86.010101} {\bibfield  {journal} {\bibinfo
  {journal} {Physical Review A}\ }\textbf {\bibinfo {volume} {86}},\ \bibinfo
  {pages} {010101} (\bibinfo {year} {2012})}\BibitemShut {NoStop}%
\bibitem [{\citenamefont {Edgar}\ \emph {et~al.}(2012)\citenamefont {Edgar},
  \citenamefont {Tasca}, \citenamefont {Izdebski}, \citenamefont {Warburton},
  \citenamefont {Leach}, \citenamefont {Agnew}, \citenamefont {Buller},
  \citenamefont {Boyd},\ and\ \citenamefont {Padgett}}]{edgar_imaging_2012}%
  \BibitemOpen
  \bibfield  {author} {\bibinfo {author} {\bibfnamefont {M.~P.}\ \bibnamefont
  {Edgar}}, \bibinfo {author} {\bibfnamefont {D.~S.}\ \bibnamefont {Tasca}},
  \bibinfo {author} {\bibfnamefont {F.}~\bibnamefont {Izdebski}}, \bibinfo
  {author} {\bibfnamefont {R.~E.}\ \bibnamefont {Warburton}}, \bibinfo {author}
  {\bibfnamefont {J.}~\bibnamefont {Leach}}, \bibinfo {author} {\bibfnamefont
  {M.}~\bibnamefont {Agnew}}, \bibinfo {author} {\bibfnamefont {G.~S.}\
  \bibnamefont {Buller}}, \bibinfo {author} {\bibfnamefont {R.~W.}\
  \bibnamefont {Boyd}},\ and\ \bibinfo {author} {\bibfnamefont {M.~J.}\
  \bibnamefont {Padgett}},\ }\bibfield  {title} {{\selectlanguage {en}\bibinfo
  {title} {Imaging high-dimensional spatial entanglement with a camera}},\
  }\href {https://doi.org/10.1038/ncomms1988} {\bibfield  {journal} {\bibinfo
  {journal} {Nature Communications}\ }\textbf {\bibinfo {volume} {3}},\
  \bibinfo {pages} {984} (\bibinfo {year} {2012})}\BibitemShut {NoStop}%
\bibitem [{\citenamefont {Dabrowski}\ \emph {et~al.}(2017)\citenamefont
  {Dabrowski}, \citenamefont {Parniak},\ and\ \citenamefont
  {Wasilewski}}]{dabrowski_einsteinpodolskyrosen_2017}%
  \BibitemOpen
  \bibfield  {author} {\bibinfo {author} {\bibfnamefont {M.}~\bibnamefont
  {Dabrowski}}, \bibinfo {author} {\bibfnamefont {M.}~\bibnamefont {Parniak}},\
  and\ \bibinfo {author} {\bibfnamefont {W.}~\bibnamefont {Wasilewski}},\
  }\bibfield  {title} {{\selectlanguage {en}\bibinfo {title}
  {Einstein–{Podolsky}–{Rosen} paradox in a hybrid bipartite system}},\
  }\href {https://doi.org/10.1364/OPTICA.4.000272} {\bibfield  {journal}
  {\bibinfo  {journal} {Optica}\ }\textbf {\bibinfo {volume} {4}},\ \bibinfo
  {pages} {272} (\bibinfo {year} {2017})}\BibitemShut {NoStop}%
\bibitem [{\citenamefont {Defienne}\ \emph
  {et~al.}(2018{\natexlab{b}})\citenamefont {Defienne}, \citenamefont
  {Reichert},\ and\ \citenamefont {Fleischer}}]{defienne_general_2018}%
  \BibitemOpen
  \bibfield  {author} {\bibinfo {author} {\bibfnamefont {H.}~\bibnamefont
  {Defienne}}, \bibinfo {author} {\bibfnamefont {M.}~\bibnamefont {Reichert}},\
  and\ \bibinfo {author} {\bibfnamefont {J.~W.}\ \bibnamefont {Fleischer}},\
  }\bibfield  {title} {\bibinfo {title} {General model of photon-pair detection
  with an image sensor},\ }\href@noop {} {\bibfield  {journal} {\bibinfo
  {journal} {Physical review letters}\ }\textbf {\bibinfo {volume} {120}},\
  \bibinfo {pages} {203604} (\bibinfo {year} {2018}{\natexlab{b}})}\BibitemShut
  {NoStop}%
\bibitem [{\citenamefont {Fedorov}\ \emph {et~al.}(2009)\citenamefont
  {Fedorov}, \citenamefont {Mikhailova},\ and\ \citenamefont
  {Volkov}}]{fedorov_gaussian_2009}%
  \BibitemOpen
  \bibfield  {author} {\bibinfo {author} {\bibfnamefont {M.~V.}\ \bibnamefont
  {Fedorov}}, \bibinfo {author} {\bibfnamefont {Y.~M.}\ \bibnamefont
  {Mikhailova}},\ and\ \bibinfo {author} {\bibfnamefont {P.~A.}\ \bibnamefont
  {Volkov}},\ }\bibfield  {title} {{\selectlanguage {en}\bibinfo {title}
  {Gaussian modelling and {Schmidt} modes of {SPDC} biphoton states}},\ }\href
  {https://doi.org/10.1088/0953-4075/42/17/175503} {\bibfield  {journal}
  {\bibinfo  {journal} {Journal of Physics B: Atomic, Molecular and Optical
  Physics}\ }\textbf {\bibinfo {volume} {42}},\ \bibinfo {pages} {175503}
  (\bibinfo {year} {2009})}\BibitemShut {NoStop}%
\bibitem [{\citenamefont {Gnatiessoro}\ \emph {et~al.}(2019)\citenamefont
  {Gnatiessoro}, \citenamefont {Mosset}, \citenamefont {Lantz},\ and\
  \citenamefont {Devaux}}]{gnatiessoro_imaging_2019-1}%
  \BibitemOpen
  \bibfield  {author} {\bibinfo {author} {\bibfnamefont {S.}~\bibnamefont
  {Gnatiessoro}}, \bibinfo {author} {\bibfnamefont {A.}~\bibnamefont {Mosset}},
  \bibinfo {author} {\bibfnamefont {E.}~\bibnamefont {Lantz}},\ and\ \bibinfo
  {author} {\bibfnamefont {F.}~\bibnamefont {Devaux}},\ }\bibfield  {title}
  {{\selectlanguage {en}\bibinfo {title} {Imaging {Spatial} {Quantum}
  {Correlations} through a thin {Scattering} {Medium}}},\ }\href
  {https://doi.org/10.1364/OSAC.2.003393} {\bibfield  {journal} {\bibinfo
  {journal} {OSA Continuum}\ }\textbf {\bibinfo {volume} {2}},\ \bibinfo
  {pages} {3393} (\bibinfo {year} {2019})}\BibitemShut {NoStop}%
\bibitem [{\citenamefont {Soro}\ \emph {et~al.}(2021)\citenamefont {Soro},
  \citenamefont {Lantz}, \citenamefont {Mosset},\ and\ \citenamefont
  {Devaux}}]{soro_quantum_2021-1}%
  \BibitemOpen
  \bibfield  {author} {\bibinfo {author} {\bibfnamefont {G.}~\bibnamefont
  {Soro}}, \bibinfo {author} {\bibfnamefont {E.}~\bibnamefont {Lantz}},
  \bibinfo {author} {\bibfnamefont {A.}~\bibnamefont {Mosset}},\ and\ \bibinfo
  {author} {\bibfnamefont {F.}~\bibnamefont {Devaux}},\ }\bibfield  {title}
  {{\selectlanguage {en}\bibinfo {title} {Quantum spatial correlations imaging
  through thick scattering media: experiments and comparison with simulations
  of the biphoton wave function}},\ }\bibfield  {journal} {\bibinfo  {journal}
  {Journal of Optics}\ }\href {https://doi.org/10.1088/2040-8986/abe1cd}
  {10.1088/2040-8986/abe1cd} (\bibinfo {year} {2021})\BibitemShut {NoStop}%
\bibitem [{\citenamefont {Morizur}\ \emph {et~al.}(2010)\citenamefont
  {Morizur}, \citenamefont {Nicholls}, \citenamefont {Jian}, \citenamefont
  {Armstrong}, \citenamefont {Treps}, \citenamefont {Hage}, \citenamefont
  {Hsu}, \citenamefont {Bowen}, \citenamefont {Janousek},\ and\ \citenamefont
  {Bachor}}]{morizur_programmable_2010}%
  \BibitemOpen
  \bibfield  {author} {\bibinfo {author} {\bibfnamefont {J.-F.}\ \bibnamefont
  {Morizur}}, \bibinfo {author} {\bibfnamefont {L.}~\bibnamefont {Nicholls}},
  \bibinfo {author} {\bibfnamefont {P.}~\bibnamefont {Jian}}, \bibinfo {author}
  {\bibfnamefont {S.}~\bibnamefont {Armstrong}}, \bibinfo {author}
  {\bibfnamefont {N.}~\bibnamefont {Treps}}, \bibinfo {author} {\bibfnamefont
  {B.}~\bibnamefont {Hage}}, \bibinfo {author} {\bibfnamefont {M.}~\bibnamefont
  {Hsu}}, \bibinfo {author} {\bibfnamefont {W.}~\bibnamefont {Bowen}}, \bibinfo
  {author} {\bibfnamefont {J.}~\bibnamefont {Janousek}},\ and\ \bibinfo
  {author} {\bibfnamefont {H.-A.}\ \bibnamefont {Bachor}},\ }\bibfield  {title}
  {{\selectlanguage {en}\bibinfo {title} {Programmable unitary spatial mode
  manipulation}},\ }\href {https://doi.org/10.1364/JOSAA.27.002524} {\bibfield
  {journal} {\bibinfo  {journal} {JOSA A}\ }\textbf {\bibinfo {volume} {27}},\
  \bibinfo {pages} {2524} (\bibinfo {year} {2010})}\BibitemShut {NoStop}%
\bibitem [{\citenamefont {Fontaine}\ \emph {et~al.}(2019)\citenamefont
  {Fontaine}, \citenamefont {Ryf}, \citenamefont {Chen}, \citenamefont
  {Neilson}, \citenamefont {Kim},\ and\ \citenamefont
  {Carpenter}}]{fontaine_laguerre-gaussian_2019}%
  \BibitemOpen
  \bibfield  {author} {\bibinfo {author} {\bibfnamefont {N.~K.}\ \bibnamefont
  {Fontaine}}, \bibinfo {author} {\bibfnamefont {R.}~\bibnamefont {Ryf}},
  \bibinfo {author} {\bibfnamefont {H.}~\bibnamefont {Chen}}, \bibinfo {author}
  {\bibfnamefont {D.~T.}\ \bibnamefont {Neilson}}, \bibinfo {author}
  {\bibfnamefont {K.}~\bibnamefont {Kim}},\ and\ \bibinfo {author}
  {\bibfnamefont {J.}~\bibnamefont {Carpenter}},\ }\bibfield  {title}
  {{\selectlanguage {en}\bibinfo {title} {Laguerre-{Gaussian} mode sorter}},\
  }\href {https://doi.org/10.1038/s41467-019-09840-4} {\bibfield  {journal}
  {\bibinfo  {journal} {Nature Communications}\ }\textbf {\bibinfo {volume}
  {10}},\ \bibinfo {pages} {1} (\bibinfo {year} {2019})}\BibitemShut {NoStop}%
\bibitem [{\citenamefont {Lib}\ \emph {et~al.}(2021)\citenamefont {Lib},
  \citenamefont {Sulimany},\ and\ \citenamefont
  {Bromberg}}]{lib_reconfigurable_2021}%
  \BibitemOpen
  \bibfield  {author} {\bibinfo {author} {\bibfnamefont {O.}~\bibnamefont
  {Lib}}, \bibinfo {author} {\bibfnamefont {K.}~\bibnamefont {Sulimany}},\ and\
  \bibinfo {author} {\bibfnamefont {Y.}~\bibnamefont {Bromberg}},\ }\bibfield
  {title} {\bibinfo {title} {Processing Entangled Photons in High Dimensions with a Programmable Light Converter},\ }\bibfield  {journal}
  {\bibinfo  {journal} {Physical Review Applied}\ }\textbf {\bibinfo {volume}
  {18(1)}},\ \bibinfo {pages} {014063} \href
  {https://journals.aps.org/prapplied/abstract/10.1103/PhysRevApplied.18.014063} {}
  (\bibinfo {year} {2022})\BibitemShut {NoStop}%
\bibitem [{\citenamefont {Brandt}\ \emph {et~al.}(2020)\citenamefont {Brandt},
  \citenamefont {Hiekkamäki}, \citenamefont {Bouchard}, \citenamefont
  {Huber},\ and\ \citenamefont {Fickler}}]{brandt_high-dimensional_2020}%
  \BibitemOpen
  \bibfield  {author} {\bibinfo {author} {\bibfnamefont {F.}~\bibnamefont
  {Brandt}}, \bibinfo {author} {\bibfnamefont {M.}~\bibnamefont {Hiekkamäki}},
  \bibinfo {author} {\bibfnamefont {F.}~\bibnamefont {Bouchard}}, \bibinfo
  {author} {\bibfnamefont {M.}~\bibnamefont {Huber}},\ and\ \bibinfo {author}
  {\bibfnamefont {R.}~\bibnamefont {Fickler}},\ }\bibfield  {title}
  {{\selectlanguage {en}\bibinfo {title} {High-dimensional quantum gates using
  full-field spatial modes of photons}},\ }\href
  {https://doi.org/10.1364/OPTICA.375875} {\bibfield  {journal} {\bibinfo
  {journal} {Optica}\ }\textbf {\bibinfo {volume} {7}},\ \bibinfo {pages} {98}
  (\bibinfo {year} {2020})}\BibitemShut {NoStop}%
\bibitem [{\citenamefont {Madonini}\ \emph {et~al.}(2021)\citenamefont
  {Madonini}, \citenamefont {Severini}, \citenamefont {Zappa},\ and\
  \citenamefont {Villa}}]{madonini_single_2021}%
  \BibitemOpen
  \bibfield  {author} {\bibinfo {author} {\bibfnamefont {F.}~\bibnamefont
  {Madonini}}, \bibinfo {author} {\bibfnamefont {F.}~\bibnamefont {Severini}},
  \bibinfo {author} {\bibfnamefont {F.}~\bibnamefont {Zappa}},\ and\ \bibinfo
  {author} {\bibfnamefont {F.}~\bibnamefont {Villa}},\ }\bibfield  {title}
  {{\selectlanguage {en}\bibinfo {title} {Single {Photon} {Avalanche} {Diode}
  {Arrays} for {Quantum} {Imaging} and {Microscopy}}},\ }\href
  {https://doi.org/10.1002/qute.202100005} {\bibfield  {journal} {\bibinfo
  {journal} {Advanced Quantum Technologies}\ }\textbf {\bibinfo {volume} {4}},\
  \bibinfo {pages} {2100005} (\bibinfo {year} {2021})}\BibitemShut {NoStop}%
\bibitem [{\citenamefont {Nomerotski}(2019)}]{nomerotski_imaging_2019}%
  \BibitemOpen
  \bibfield  {author} {\bibinfo {author} {\bibfnamefont {A.}~\bibnamefont
  {Nomerotski}},\ }\bibfield  {title} {{\selectlanguage {en}\bibinfo {title}
  {Imaging and time stamping of photons with nanosecond resolution in {Timepix}
  based optical cameras}},\ }\href {https://doi.org/10.1016/j.nima.2019.05.034}
  {\bibfield  {journal} {\bibinfo  {journal} {Nuclear Instruments and Methods
  in Physics Research Section A: Accelerators, Spectrometers, Detectors and
  Associated Equipment}\ }\textbf {\bibinfo {volume} {937}},\ \bibinfo {pages}
  {26} (\bibinfo {year} {2019})}\BibitemShut {NoStop}%
\bibitem [{\citenamefont {Schneeloch}\ and\ \citenamefont
  {Howell}(2016)}]{schneeloch_introduction_2016}%
  \BibitemOpen
  \bibfield  {author} {\bibinfo {author} {\bibfnamefont {J.}~\bibnamefont
  {Schneeloch}}\ and\ \bibinfo {author} {\bibfnamefont {J.~C.}\ \bibnamefont
  {Howell}},\ }\bibfield  {title} {{\selectlanguage {en}\bibinfo {title}
  {Introduction to the transverse spatial correlations in spontaneous
  parametric down-conversion through the biphoton birth zone}},\ }\href
  {https://doi.org/10.1088/2040-8978/18/5/053501} {\bibfield  {journal}
  {\bibinfo  {journal} {Journal of Optics}\ }\textbf {\bibinfo {volume} {18}},\
  \bibinfo {pages} {053501} (\bibinfo {year} {2016})}\BibitemShut {NoStop}%
\bibitem [{\citenamefont {Rech}\ \emph {et~al.}(2008)\citenamefont {Rech},
  \citenamefont {Ingargiola}, \citenamefont {Spinelli}, \citenamefont
  {Labanca}, \citenamefont {Marangoni}, \citenamefont {Ghioni},\ and\
  \citenamefont {Cova}}]{rech_optical_2008}%
  \BibitemOpen
  \bibfield  {author} {\bibinfo {author} {\bibfnamefont {I.}~\bibnamefont
  {Rech}}, \bibinfo {author} {\bibfnamefont {A.}~\bibnamefont {Ingargiola}},
  \bibinfo {author} {\bibfnamefont {R.}~\bibnamefont {Spinelli}}, \bibinfo
  {author} {\bibfnamefont {I.}~\bibnamefont {Labanca}}, \bibinfo {author}
  {\bibfnamefont {S.}~\bibnamefont {Marangoni}}, \bibinfo {author}
  {\bibfnamefont {M.}~\bibnamefont {Ghioni}},\ and\ \bibinfo {author}
  {\bibfnamefont {S.}~\bibnamefont {Cova}},\ }\bibfield  {title}
  {{\selectlanguage {en}\bibinfo {title} {Optical crosstalk in single photon
  avalanche diode arrays: a new complete model}},\ }\href
  {https://doi.org/10.1364/OE.16.008381} {\bibfield  {journal} {\bibinfo
  {journal} {Optics Express}\ }\textbf {\bibinfo {volume} {16}},\ \bibinfo
  {pages} {8381} (\bibinfo {year} {2008})}\BibitemShut {NoStop}%
\bibitem [{\citenamefont {Connolly}\ \emph {et~al.}(2019)\citenamefont
  {Connolly}, \citenamefont {Ren}, \citenamefont {Henderson},\ and\
  \citenamefont {Buller}}]{connolly_hot_2019}%
  \BibitemOpen
  \bibfield  {author} {\bibinfo {author} {\bibfnamefont {P.}~\bibnamefont
  {Connolly}}, \bibinfo {author} {\bibfnamefont {X.}~\bibnamefont {Ren}},
  \bibinfo {author} {\bibfnamefont {R.}~\bibnamefont {Henderson}},\ and\
  \bibinfo {author} {\bibfnamefont {G.}~\bibnamefont {Buller}},\ }\bibfield
  {title} {{\selectlanguage {en}\bibinfo {title} {Hot pixel classification of
  single-photon avalanche diode detector arrays using a log-normal statistical
  distribution}},\ }\href {https://doi.org/10.1049/el.2019.1427} {\bibfield
  {journal} {\bibinfo  {journal} {Electronics Letters}\ }\textbf {\bibinfo
  {volume} {55}},\ \bibinfo {pages} {1004} (\bibinfo {year}
  {2019})}\BibitemShut {NoStop}%
\bibitem [{\citenamefont {Eckmann}\ \emph {et~al.}(2020)\citenamefont
  {Eckmann}, \citenamefont {Bessire}, \citenamefont {Unternährer},
  \citenamefont {Gasparini}, \citenamefont {Perenzoni},\ and\ \citenamefont
  {Stefanov}}]{eckmann_characterization_2020-2}%
  \BibitemOpen
  \bibfield  {author} {\bibinfo {author} {\bibfnamefont {B.}~\bibnamefont
  {Eckmann}}, \bibinfo {author} {\bibfnamefont {B.}~\bibnamefont {Bessire}},
  \bibinfo {author} {\bibfnamefont {M.}~\bibnamefont {Unternährer}}, \bibinfo
  {author} {\bibfnamefont {L.}~\bibnamefont {Gasparini}}, \bibinfo {author}
  {\bibfnamefont {M.}~\bibnamefont {Perenzoni}},\ and\ \bibinfo {author}
  {\bibfnamefont {A.}~\bibnamefont {Stefanov}},\ }\bibfield  {title}
  {{\selectlanguage {en}\bibinfo {title} {Characterization of space-momentum
  entangled photons with a time resolving {CMOS} {SPAD} array}},\ }\href
  {https://doi.org/10.1364/OE.401260} {\bibfield  {journal} {\bibinfo
  {journal} {Optics Express}\ }\textbf {\bibinfo {volume} {28}},\ \bibinfo
  {pages} {31553} (\bibinfo {year} {2020})},\ \bibinfo {note} {publisher:
  Optical Society of America}\BibitemShut {NoStop}%
\bibitem [{\citenamefont {Tasca}\ \emph {et~al.}(2018)\citenamefont {Tasca},
  \citenamefont {Sánchez}, \citenamefont {Walborn},\ and\ \citenamefont
  {Rudnicki}}]{tasca_mutual_2018}%
  \BibitemOpen
  \bibfield  {author} {\bibinfo {author} {\bibfnamefont {D.~S.}\ \bibnamefont
  {Tasca}}, \bibinfo {author} {\bibfnamefont {P.}~\bibnamefont {Sánchez}},
  \bibinfo {author} {\bibfnamefont {S.~P.}\ \bibnamefont {Walborn}},\ and\
  \bibinfo {author} {\bibfnamefont {E.}~\bibnamefont {Rudnicki}},\ }\bibfield
  {title} {\bibinfo {title} {Mutual {Unbiasedness} in {Coarse}-{Grained}
  {Continuous} {Variables}},\ }\href
  {https://doi.org/10.1103/PhysRevLett.120.040403} {\bibfield  {journal}
  {\bibinfo  {journal} {Physical Review Letters}\ }\textbf {\bibinfo {volume}
  {120}},\ \bibinfo {pages} {040403} (\bibinfo {year} {2018})}\BibitemShut
  {NoStop}%
\bibitem [{\citenamefont {Abouraddy}\ \emph {et~al.}(2002)\citenamefont
  {Abouraddy}, \citenamefont {Saleh}, \citenamefont {Sergienko},\ and\
  \citenamefont {Teich}}]{abouraddy_entangled-photon_2002-2}%
  \BibitemOpen
  \bibfield  {author} {\bibinfo {author} {\bibfnamefont {A.~F.}\ \bibnamefont
  {Abouraddy}}, \bibinfo {author} {\bibfnamefont {B.~E.~A.}\ \bibnamefont
  {Saleh}}, \bibinfo {author} {\bibfnamefont {A.~V.}\ \bibnamefont
  {Sergienko}},\ and\ \bibinfo {author} {\bibfnamefont {M.~C.}\ \bibnamefont
  {Teich}},\ }\bibfield  {title} {{\selectlanguage {en}\bibinfo {title}
  {Entangled-photon {Fourier} optics}},\ }\href
  {https://doi.org/10.1364/JOSAB.19.001174} {\bibfield  {journal} {\bibinfo
  {journal} {JOSA B}\ }\textbf {\bibinfo {volume} {19}},\ \bibinfo {pages}
  {1174} (\bibinfo {year} {2002})}\BibitemShut {NoStop}%
\bibitem [{\citenamefont {Reichert}\ \emph {et~al.}(2018)\citenamefont
  {Reichert}, \citenamefont {Defienne},\ and\ \citenamefont
  {Fleischer}}]{reichert_optimizing_2018}%
  \BibitemOpen
  \bibfield  {author} {\bibinfo {author} {\bibfnamefont {M.}~\bibnamefont
  {Reichert}}, \bibinfo {author} {\bibfnamefont {H.}~\bibnamefont {Defienne}},\
  and\ \bibinfo {author} {\bibfnamefont {J.~W.}\ \bibnamefont {Fleischer}},\
  }\bibfield  {title} {\bibinfo {title} {Optimizing the signal-to-noise ratio
  of biphoton distribution measurements},\ }\href
  {https://doi.org/10.1103/PhysRevA.98.013841} {\bibfield  {journal} {\bibinfo
  {journal} {Physical Review A}\ }\textbf {\bibinfo {volume} {98}},\ \bibinfo
  {pages} {013841} (\bibinfo {year} {2018})}\BibitemShut {NoStop}%
\end{thebibliography}

%
%
%
\end{document}